\documentclass[12pt]{article}
\usepackage{graphicx,floatflt,amssymb,epsfig,epsf} 
 \usepackage{ulem}
\def\gtap{\raisebox{-.4ex}{\rlap{$\sim$}} \raisebox{.4ex}{$>$}}

\def\be {\begin{equation}}
\def\ee {\end{equation}}
\def\bea {\begin{eqnarray}}
\def\eea {\end{eqnarray}}

\def\opcit(#1){ {\em op. cit.}, #1}

\def\issue(#1,#2,#3){#1 (#3) #2} 

\def\APP(#1,#2,#3){Acta Phys.\ Polon.\ \issue(#1,#2,#3)}
\def\ARNPS(#1,#2,#3){Ann.\ Rev.\ Nucl.\ Part.\ Sci.\ \issue(#1,#2,#3)}
\def\CPC(#1,#2,#3){Comp.\ Phys.\ Comm.\ \issue(#1,#2,#3)}
\def\CIP(#1,#2,#3){Comput.\ Phys.\ \issue(#1,#2,#3)}
\def\EPJC(#1,#2,#3){Eur.\ Phys.\ J.\ C\ \issue(#1,#2,#3)}
\def\EPJD(#1,#2,#3){Eur.\ Phys.\ J. Direct\ C\ \issue(#1,#2,#3)}
\def\IEEETNS(#1,#2,#3){IEEE Trans.\ Nucl.\ Sci.\ \issue(#1,#2,#3)}
\def\IJMP(#1,#2,#3){Int.\ J.\ Mod.\ Phys. \issue(#1,#2,#3)}
\def\JHEP(#1,#2,#3){J.\ High Energy Physics \issue(#1,#2,#3)}
\def\JPG(#1,#2,#3){J.\ Phys.\ G \issue(#1,#2,#3)}
\def\MPL(#1,#2,#3){Mod.\ Phys.\ Lett.\ \issue(#1,#2,#3)}
\def\NP(#1,#2,#3){Nucl.\ Phys.\ \issue(#1,#2,#3)}
\def\NIM(#1,#2,#3){Nucl.\ Instrum.\ Meth.\ \issue(#1,#2,#3)}
\def\PL(#1,#2,#3){Phys.\ Lett.\ \issue(#1,#2,#3)}
\def\PRD(#1,#2,#3){Phys.\ Rev.\ D \issue(#1,#2,#3)}
\def\PRL(#1,#2,#3){Phys.\ Rev.\ Lett.\ \issue(#1,#2,#3)}
\def\SJNP(#1,#2,#3){Sov.\ J. Nucl.\ Phys.\ \issue(#1,#2,#3)}
\def\ZPC(#1,#2,#3){Zeit.\ Phys.\ C \issue(#1,#2,#3)}

\newcommand\npb[3]{{Nucl.\ Phys.\ }{\bf B #1} (#2) #3}
\newcommand\zpc[3]{{Z.\ Physik }{\bf C #1} (#2) #3}

\begin{document} 
\begin{flushright} 
TIFR/TH/10-14\\
RECAPP-HRI-2010-009\\
\end{flushright} 
\vskip 30pt 
 
\begin{center} 
{\Large \bf Search for the minimal universal extra dimension model at the LHC with $\sqrt{s}$=7 TeV}\\
\vspace*{1cm} 
\renewcommand{\thefootnote}{\fnsymbol{footnote}} 
{\large {\sf Biplob Bhattacherjee$^{a}$} and {\sf Kirtiman Ghosh$^{b}$} } \\ 
\vspace{10pt}
\bigskip
 $^a${\footnotesize\rm 
Department of Theoretical Physics, Tata Institute of Fundamental Research, \\ 
1, Homi Bhabha Road, Mumbai 400 005, India. \\
E-mail: {\sf biplob@theory.tifr.res.in, } }

\bigskip
$^b${\footnotesize\rm 
 Harish Chanda Research Institute, Chhatnag Road, \\
Jhunsi, Allahabad 211019, India \\
 E-mail: {\sf kirtiman@hri.res.in. }}

\normalsize 
\end{center} 

\begin{abstract}
\noindent Universal Extra Dimension (UED) model is one of the popular extension of the Standard Model (SM) 
which offers interesting phenomenology. In the minimal UED (mUED) model, Kaluza-Klein (KK)
parity conservation ensures that $n=1$ KK states can only be pair produced at colliders and the lightest 
KK particle is stable. In most of the parameter space, first KK excitation of SM hypercharge gauge 
boson is the lightest one and it can be a viable dark matter candidate. Thus, the decay of $n=1$ KK 
particles will always involve missing transverse energy ($E_T\!\!\!\!\!\!/~$) as well as leptons and jets. The production cross 
sections of $n=1$ KK particles are large and such particles may be observed at the Large Hadron Collider
 (LHC). We explore the mUED discovery potential of the LHC with $\sqrt{s}$ = 7 TeV 
in the  multileptonic final states. Since in the early LHC run, precise determination of $E_T\!\!\!\!\!\!/~~$ may not be possible,
 we examine the LHC reach 
with and without using $E_T\!\!\!\!\!\!/~$ information. We observe that $E_T\!\!\!\!\!\!/~$ cut 
will not improve mUED discovery reach significantly. We have found that opposite sign di-lepton channel is 
the most promising discovery mode and with first $fb^{-1}$ of collected luminosity, LHC will be able to 
discover the strongly interacting $n=1$ KK particles with masses upto 800 $\sim$ 900 GeV.

\vskip 5pt \noindent 
\texttt{PACS numbers:~ 13.85.Qk, 12.60.-i,14.80.Rt} \\ 
\texttt{Keywords:~~Universal Extra Dimension, LHC, Multilepton search}
\end{abstract}

\renewcommand{\thesection}{\Roman{section}} 
\setcounter{footnote}{0} 
\renewcommand{\thefootnote}{\arabic{footnote}} 
\newpage
\section{Introduction}
\noindent The Universal Extra Dimension (UED) model, 
proposed by Appelquist, Cheng and Dobrescu 
\cite{Appelquist:2000nn},
 appears to be one of the popular scenario 
beyond the Standard Model (SM). This model 
assumes that all particles can propagate in 
the flat extra dimensions and in its minimal version (mUED) \cite{mued1,mued2}, there is only 
one extra dimension $y$ compactified on a 
circle of radius $R$ ($S_1$ symmetry).
An additional $Z_2$ symmetry, which identifies 
$y$ to $-y$ is required to get zero mode chiral 
fermions at low energy. 
The $Z_2$ symmetry breaks the translational 
invariance along the $5$th dimension and generates two 
fixed points at $y=0$ and $y=\pi R$. The size of 
the extra dimension is taken to be small enough 
so that one can dimensionally reduce the theory and 
construct the effective 4D Lagrangian. The low energy 
effective  
Lagrangian contains infinite number of Kaluza-Klein (KK) excitations 
(identified by an integer number n, called the KK number)
for all the fields which are present in the 
higher dimensional Lagrangian. The zero modes of 
the KK towers are generally identified with the 
Standard Model particles. 

On the other hand, there is a particular variant of the UED model, called 
two Universal Extra Dimension (2UED) model \cite{2ued}, 
where all the SM fields propagate in $(5+1)$ dimensional space time. 
2UED model has some additional attractive
features. As an example, 2UED model can naturally explain the long life time of
proton \cite{dobrescu} and more interestingly it predicts that the
number of fermion generations should be an integral multiple of three
\cite{dobrescu1}. 
The phenomenology of KK-excitations 
of 2UED model has been studied in great detail in recent times 
which covers its implications at colliders \cite{coll2ued}, in various
low-energy observables \cite{low2ued} and dark
matter/cosmology \cite{dark2ued}. However, in this article, we will 
only concentrate on the phenomenology of mUED model at the Large Hadron Collider (LHC).

\noindent One of the interesting feature of the mUED model is the conservation of KK number.
Since all particles can propagate in the extra dimension, the momentum along the extra dimension 
is conserved and it is also quantized because of the compactification of the extra dimension $y$. 
The five dimensional momentum conservation is translated into the conservation of KK number 
in the four dimensions.
 However, the presence of  
two fixed points break the translational symmetry and KK number is not a good quantum number. 
In principle, there may exist some 
operators located at these fixed points and 
one can expect mixing among different KK states. However, if the localized operators are 
symmetric under the exchange\footnote{This is another $Z_2$ symmetry, but 
not the $Z_2$ of $y \leftrightarrow$ $-y$.} of the fixed points 
, the conservation of KK number breaks down to the conservation 
of KK parity defined as $(- 1)^n$, where $n$ is the KK number. 
The conservation of KK parity ensures that $n=1$ particles are always produced in pairs 
and the lightest $n=1$ particle (LKP) must be stable. It also forbids tree level mUED contribution to 
any SM process. The situation is analogous
to the $R$ parity conserving supersymmetric models \cite{susy}. The tree level mUED spectrum is extremely 
degenerate and first excitation of any massless SM particle can be the LKP (say KK excitation of gluon, 
photon, neutrino i.e., $g_1$, $\gamma_1$,
$\nu_1$ etc.). The radiative corrections \cite{mued1,georgi,rad1} play very important role in determining the
mUED mass spectrum. The 
correction terms can be finite (bulk correction) or it may depends on the cut-off of the model $\Lambda$
(boundary correction). The cut-off dependence comes from the fact that mUED is a higher dimensional model 
and thus, it is nonrenormalizable. This model should be treated in the spirit of an effective theory 
valid upto a scale $\Lambda > R^{-1}$. It is shown in the literature \cite{mued1} that radiative correction partially removes the 
degeneracy in the spectrum and in most of the parameter space $n=1$ excitation of hypercharge gauge 
boson $B$ called $\gamma_1$ \footnote {Actually in the presence of radiative correction, the KK 
Weinberg angle is small so that $B_1$ $\approx$ $\gamma_1$ and $W^{3}_1$ $\approx$ $Z_1$. }, is the LKP. 
The $\gamma_1$ can produce 
the right amount of cosmological relic density (consistent with the WMAP data) and turns out to 
be a good dark matter candidate \cite{darkued}. The mass of LKP is approximately $R^{-1}$ and 
hence, the overclosure of the
 universe puts an upper bound on $R^{-1} < $ 1400 GeV. The lower limit of $R^{-1}$ 
comes from the low energy observables and direct search of new particles at the Tevatron. 
Constraints on $R^{-1}$ from $g-2$ of the muon \cite{g_muon}, flavour changing
neutral currents \cite{chk,buras,desh}, $Z \to b\bar{b}$ decay
\cite{santa}, the $\rho$ parameter \cite{Appelquist:2000nn,appel-yee}, other
electroweak precision tests \cite{ewued}, hadron
collider studies \cite{Abbott,lin} imply that $R^{-1}~\gtap~300$
GeV. A recent inclusive $\bar{B} \to X_s \gamma$ analysis sets a
stronger constraint $R^{-1}~\gtap~600$ GeV at 95$\%$ CL \cite{Haisch:2007vb}, although it
still keeps open a lower value of $R^{-1}$ $\sim$ 400 GeV at 99$\%$ CL. The lower and upper bounds indicate that atleast pair production of  
$n = 1$ excited states are possible at the LHC. The masses of KK particles
are dependent on $\Lambda$ which has no determined value. One loop corrected
$SU(3)$, $SU(2)$ and $U(1)$ gauge couplings show power law running in the mUED model and  
almost meet at the scale $\Lambda$= 20 $R^{-1}$ \cite{dienes,Bhattacharyya}. Thus, one can take 
$\Lambda  =20 R^{-1}$ as the 
cut-off of the model and expect the presence of some new physics above that energy scale. 
If we neglect such type of unification, 
we can extend the value of the cut-off, but at the scale around 40 $R^{-1}$, $U(1)$ coupling becomes 
nonperturbative. Thus, the cut-off  $\Lambda R$ of the mUED model should not be taken above 40 and
one should use the value of $\Lambda R$ between $10$ to 40.  \\

The collider phenomenology of mUED model is similar to the $R$ parity conserving supersymmetry
  and it has been investigated in Ref. \cite{collider}, though
not with the same level of detail and sophistication as the corresponding signals from
supersymmetry \cite{susycoll,Baer:2010tk}. 
The crucial feature of these studies is the 
existence of the stable LKP (KK parity conservation) and hence, the missing energy and missing transverse 
momentum signal at the colliders. The current bounds allow KK states 
to be as light as a few hundreds of GeV. Such light $n=1$ particles can be pair produced at 
the LHC and they cascade down to $\gamma_1$ which escapes the detector. Like $R_p$ conserving SUSY, 
the characteristic signature of mUED is leptons plus jets accompanied by missing energy. The SM 
particles arising from the decay of KK particles are usually soft. It is due to the partial degeneracy 
in the mUED spectrum\footnote{This feature uncommon in most of the supersymmetric models  and one can use this  
to discriminate mUED from different supersymmetric scenarios \cite{Bhattacherjee:2009jh}.}. The model predictions depend
only on the two parameters $R^{-1}$ and $\Lambda$. \\
 
One of the cleanest signal for new physics search is the multiple leptons in the final 
state. Leptonic signals at the LHC with missing energy have been studied in the literature – 
particularly in the context of SUSY searches. 
In the present work, we also consider leptonic final states and systematically study the 
capability of LHC with $\sqrt s=7$ TeV in the discovery of mUED. 
We consider two\footnote {We do not consider single lepton channel 
here. This is because, SM background coming from the $W$
production is huge and can not be eliminated by choosing proper cuts.} to four leptons accompanied 
by jets and missing transverse energy in the final states. There are two motivations for our present 
study. First, most of the phenomenological study on mUED was carried out using parton level 
monte carlo simulations and a more detailed analysis is required at this stage. 
The other reason is that people have studied
 mUED model by assuming centre of mass energy of LHC to be equal to 14 TeV but, LHC has started its 
operation with reduced CM energy of 7 TeV and it will collect data upto 1fb$^{-1}$. 
It is therefore, useful to recalculate 
the mUED discovery reach of LHC with CM energy $7$ TeV.
 The rest of the paper is arranged as follows. In the next section, we shall 
discuss the production of multileptonic final states in mUED model and comment on possible 
SM backgrounds. In section III, we shall define the final states, event selection criteria and explore the 
discovery reach of the LHC. Finally, section IV summarizes our main results and addresses the possible 
issues of this work.

\section{Leptonic signals of mUED model}

Leptonic final states are always represented as the most powerful discovery channels at the hadron collider. 
The detectors of LHC have the ability to identify the leptons very efficiently 
within a very wide energy range.   
In mUED model, leptonic branching ratios are always favorable and  multilepton signals can become 
competitive to the SM backgrounds. 

At the LHC, the dominant production processes of mUED are the pair 
production of $n=1$ KK quarks and KK gluons. Typical mUED spectrum shows that the colored KK states are 
heavier than the electroweak KK particles and $n = 1$ gluon $g_1$ is the heaviest. 
It can decay to both 
$n=1$ singlet ($q_1$) and doublet ($Q_1$) quarks with almost same branching ratios, 
although, there is a slight kinematic 
preference to the singlet channel. The singlet quark can decay only to $\gamma_1$ and SM quark. 
On the other hand, doublet quarks decay mostly to  $W_1$ or $Z_1$ (which are KK excitation of 
electroweak gauge boson $W$ and $Z$). Hadronic decay modes of 
$W_1$ and $Z_1$ are closed kinematically and these can decay universally to all $n=1$ doublet lepton 
flavours ($L_1$ or $\nu_1$). 
$W_1$ decays to $L_1^{\pm} \nu_{L}$ or $L^{\pm} \nu_{1}$ with equal branching ratios. Similarly, 
$Z_1$ can decay only to $L_1 l$ or $\nu_1 {\nu}$. The KK leptons finally decay to $\gamma_1$ 
and a ordinary (SM) lepton. The $\gamma_1$ will escape the detector because it has no 
strong and electromagnetic interactions. Thus, the resulting signature would then be $n$ jets + $m$ leptons + 
missing $E_T$. The pair production of $n=1$ singlet quarks ($q_1$) lead to 2 jets plus missing energy signature. 
Multijet with missing energy signal has become a canonical signature for SUSY search. To reduce the SM background contributions, all such 
analysis demand that the missing transverse energy must be greater than 200 GeV. 
In the next section, we shall show that the $E_T\!\!\!\!\!\!/~~$ distribution of mUED models is softer and thus, such type 
of cut will kill the SM backgrounds as well as mUED signal. Therefore, the more interesting processes that we 
can consider are $g_1 g_1$, $g_1 Q_1$ or $Q_1 Q_1$, where $Q_1$ is the $n=1$ doublet KK quarks. In this case, the leptons are 
produced from the cascade decay of $W_1$ and $Z_1$. The number of lepton may vary from $1$ to $4$. 
At this point, let us comment on 
third generation KK states. The singlet and doublet KK states are not exactly the mass 
eigenstates. In the doublet-singlet basis, mass matrix is non-diagonal, and the off-diagonal elements 
are the zeroth level mass $M_0$. We can neglect the SM masses of the quarks and leptons with respect to 
$R^{-1}$ except for the third generation quarks. After diagonalisation 
and a chiral rotation, one gets the proper mass eigenstates called $\tilde{q_1}$ and $\tilde{q_2}$. 
In case of KK top quarks, the doublet singlet mixing is most prominent. $\tilde{t_1}$ can decay 
dominantly to charged KK Higgs, and  $\tilde{t_2}$ can decay to $W_1$ or $Z_1$. Generally the KK 
top quarks are taken separately in the analysis and it is not included in our results. The decay of 
doublet b KK quark to SM top and $W_1$ is not allowed kinematically and its main decay mode is SM $b$ 
quark with $n=1$ Z boson. \\

At the LHC, one can also have associated production 
of a $n = 1$ KK electroweak gauge boson together with a $n = 1$ KK quark. Such processes also 
contribute to the multileptonic final states with low jet multiplicity, but the cross sections are
much lower than the strong cross sections. As an example, $Q_1 W_1$ production cross 
section is around 5 pb for $R^{-1}$=300 GeV and $\Lambda R$=20 at LHC with centre of mass energy 14 TeV,
whereas strong KK pair production is around 2000 pb. Similarly, the production and decays of the 
$n=2$ particles result in final states with leptons. Inclusion of all such processes will definitely 
increase the mUED discovery reach at the LHC, but our analysis is restricted to the pair production of 
strongly interacting KK particles only.

The KK quarks and gluons carry colours and it is needless to mention that their production cross 
sections are high at the LHC. The tree level KK number conserving couplings of KK quarks and gluons are 
similar to SM couplings and there is no $\Lambda$ or $R$ dependence in their couplings. 
However, the masses of the KK states are logarithmically cut-off dependent and thus, the KK production 
cross sections depend mildly on the cut-off. Here we fix the value of $\Lambda R$ equals to 20 and 
plot the the leading order pair production cross sections of coloured $n=1$ KK states as a function of $R^{-1}$ 
at the  LHC with $\sqrt{s}$= $7$ TeV. For completeness, we also plot the same for LHC with 
$\sqrt{s}$=14 TeV. We use CTEQ5L parton distribution function and the scale of the strong 
coupling constant is taken to be equal to the parton level center of mass energy. 
Here we sum over the final state quark flavors and include charge-conjugated contributions. \\

\begin{figure}
\begin{center}
\epsfig{figure=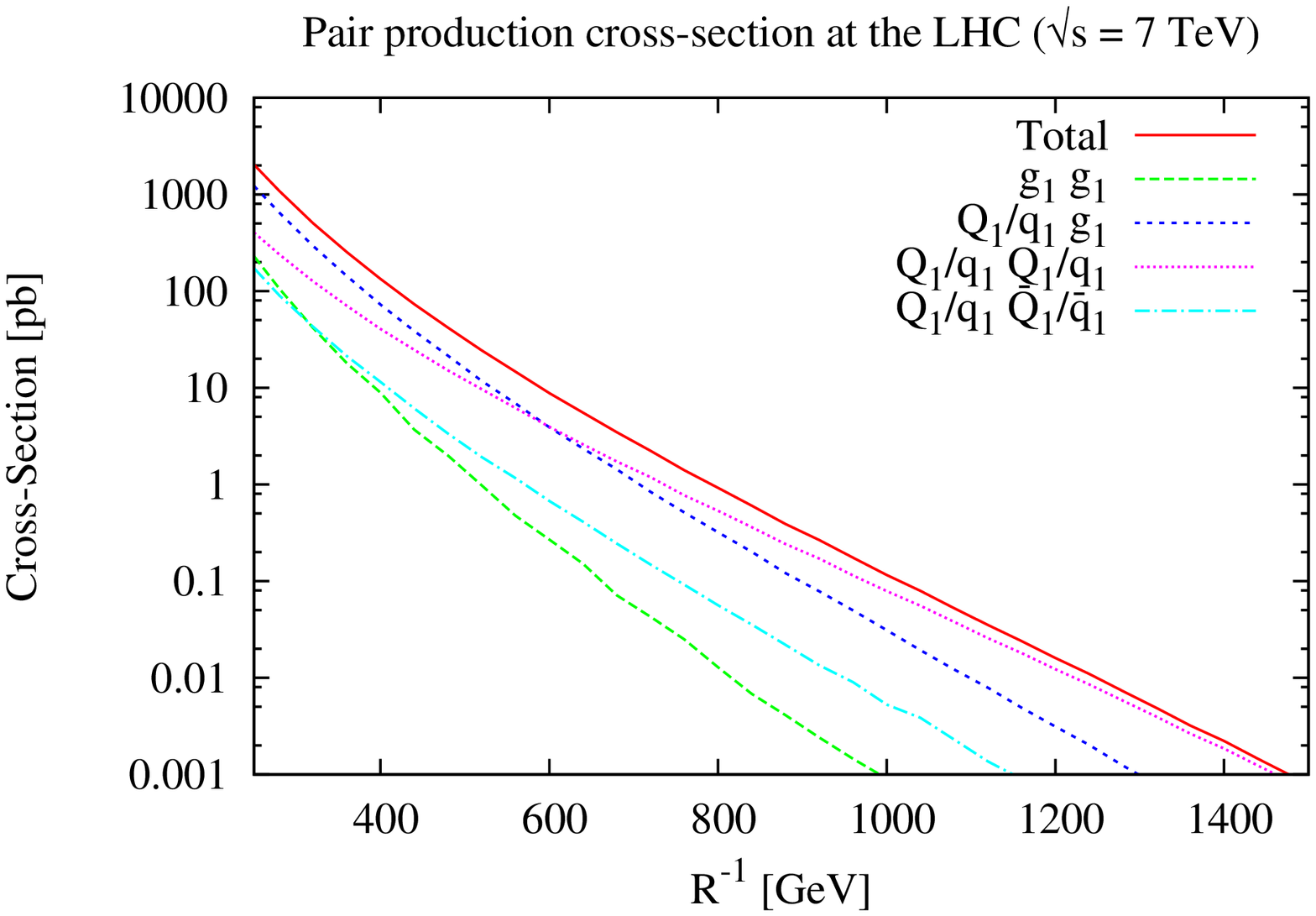,height=12cm,width=6.5cm,angle=-90}\\
\epsfig{figure=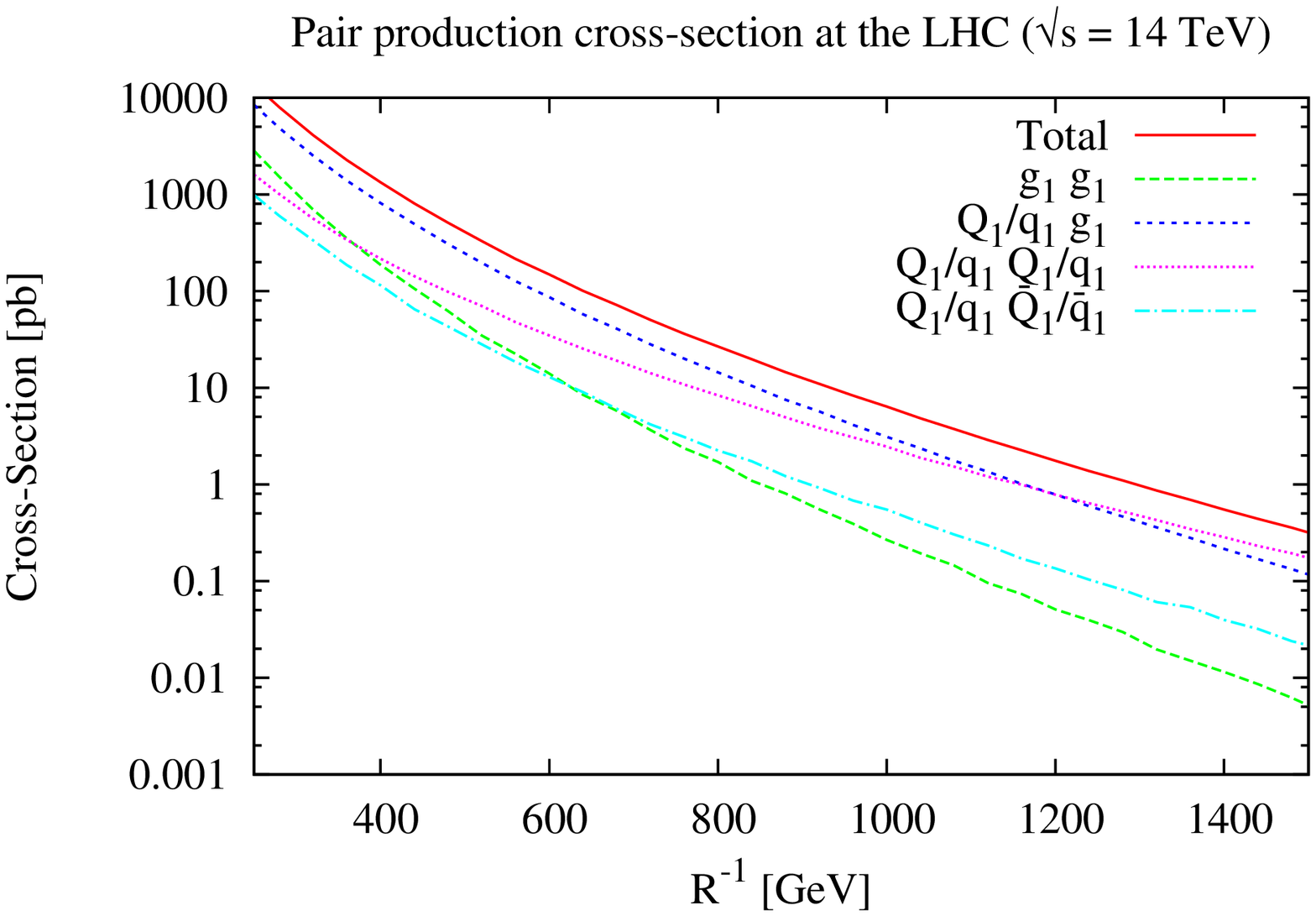,height=12cm,width=6.5cm,angle=-90}
 \caption{Cross sections for the pair production of n=1 quarks and gluons at the LHC as a function
of $R^{-1}$. $\Lambda R$ is kept fixed at 20. The top figure is for center of mass energy 7 TeV whereas the bottom
is for center of mass energy 14 TeV. }
 \label{fig:cross_7_14}
\end{center}
\end{figure}

From Fig. \ref{fig:cross_7_14}, we see that at the 7 TeV LHC, if $R^{-1}$ is less than 600 GeV, 
$q_1/Q_1~ g_1$ cross section is dominant and the cross sections fall off rapidly with $R^{-1}$. 
The cascade decay of $Q_1~g_1$ events generally produce 3 jets, leptons and missing energy.   
At the LHC with $\sqrt s=7$ TeV, the total mUED pair production cross section is about 100 pb
for $R^{-1}$=300 GeV and it is around 1pb for $R^{-1}$=800 GeV corresponding to $10^{5}$ to $10^{3}$ 
events with $1 fb^{-1}$ of luminosity. From this naive result, we may say that even with small amount
of data, we can get some hints of KK particles. The reach of the 14 TeV LHC is much higher and a 
run with moderate luminosity has the capability to probe the entire dark matter allowed region of 
mUED parameter space. Again in case of 14 TeV LHC, $Q_1 g_1$ rate is the highest upto $R^{-1}$=1200 GeV,
above which $n=1$ quark pair production is comparable. It is to be noted that we have used
the leading order cross sections from PYTHIA \cite{PYTHIA}, but the higher order 
correction to the mUED processes should be significant. \\

We are now in a position to specify the final states used in our analysis. 
In this part, we shall briefly describe how one can get di-leptons, 
tri-leptons and four leptons associated with jets and missing 
transverse momentum in the mUED model. We shall also comment on the 
possible SM backgrounds. In the SM, leptonic final states arise  from the decay of 
$W/Z$ boson, top quark or from the semileptonic decay of heavy flavours.  
The Standard Model background processes most relevant to mUED searches are $t \bar{t}$,
$W$ plus jets, $Z$ plus jets, diboson production etc. A good understanding of such 
processes are essential and precise predictions provide a high sensitivity to 
searches for new phenomena producing multi-leptonic final states.\\

{\bf \underline {$2 l$ final state}}: We look for an excess over the SM background in a 
final state that contains exactly equal to two leptons ($e$ or $\mu$) plus jets and significant amount 
of missing energy. The leptons can have same or opposite charges. In the mUED model, the opposite sign 
di-leptons can be produced from the production of opposite sign doublet quarks (like $U_1$ $\bar{U}_1$ production) 
and its decay through $W_1^{\pm}$. The each $W_1$  can produce one lepton and these two leptons have  
opposite charges. The production of $g_1 Q_1$ and $g_1 g_1$ followed by the 2 body decay of KK gluon to doublet KK quarks 
can also produce two opposite sign doublet quarks. 
KK $Z$ boson is another source of di-leptons. The same sign di-lepton 
production mechanism is similar to the opposite sign case. In this case, two same sign doublet quark is 
needed which can be produced directly or from the $g_1 Q_1$ and $g_1 g_1$ processes. The same sign doublet 
quark pair production via the exchange of KK gauge bosons (mainly $g_1$) is the most significant 
production mechanism. One can 
easily estimate the number of di-leptons in the final state from the knowledge of cross sections, because 
the branchings are quite insensitive to the model parameters of the model. The OSD final state 
has a higher rate than the SSD final state.    
The advantage of SSD over OSD signal is that it is almost background free. The  most 
significant SM background of same sign di-lepton is coming from $t \bar{t}$ production followed 
by the semileptonic decay of $t \bar t$ pairs. Here the second lepton originates from the semileptonic $b$ decay. 
The $b \bar{b}$ production can also be a dangerous background where 
the both leptons may come from $b$ decay. The leptons from the decay of $b$ quarks are generally soft 
and non isolated from the associated jets. Therefore, by requiring high $p_T$ leptons and tight 
isolation criteria one can suppress such background. In addition, diboson production also contributes 
to the background. In real experiment, charge mismeasurement of lepton is also possible. 
This can turn an opposite-sign pair into a like-sign pair.  Estimate of the sign misidentification 
rate is necessary and it should be included in the study. But, we have not done any detector 
simulation and hence, charge misidentification is neglected. In case of opposite sign di-lepton final state, 
$Z$ boson, $t \bar{t}$ and diboson production are dominant backgrounds. By putting invariant
mass cut on the di-leptons, Z boson background can be reduced, but $t \bar{t}$ background is irreducible. 
In mUED model, the leptons and jets are generally soft and the missing transverse energy distribution is also 
soft. So, it is not possible to reduce  SM $t \bar{t}$ background by applying hard cuts on the  
final state leptons, jets or missing $E_T$. \\

{\bf \underline {$3 l$ final state}} Next we consider tri-leptonic final state. Isolated tri-lepton 
 signal comes from the cascade production of $W_1 Z_1$ gauge boson pairs and their leptonic decays. 
In SUSY, first chargino and second neutralino are lighter than the gluino and
squarks, and typically masses of these particles lie in the range 200-300 GeV. Therefore, chargino neutralino 
production rate is expected to be high at the LHC. The production of neutralino chargino pair leads to 
the hadronically quiet tri-lepton signal which is treated as the golden discovery channel for SUSY. 
In case of mUED, the masses of the $W_1$ or $Z_1$ are $\sim$ $R^{-1}$, therefore, $W_1 Z_1$ direct pair production 
cross section is much smaller than the SUSY chargino neutrlino production rate. For this reason, 
one has to consider the cascade decay of KK coloured states to electroweak gauge bosons. Thus, 
in case of mUED, tri-lepton signal is usually associated with jets. The of jets coming from the decay 
of KK quarks and gluons generally soft and a small fraction of the events may be hadronically quiet.
Several SM processes 
can mimic the tri-lepton signature. The main backgrounds to the tri-leptons are from 
$t \bar{t}$, $t \bar t Z,~ t \bar t W$, $ZZ$, $WZ$ etc. The backgrounds containing an real $Z$-boson can be rejected by choosing 
the invariant mass of the same flavour oppositely charged lepton to be less than 80 GeV or more than
100 GeV.

{\bf \underline {$4 l$ final state}} The four lepton signals originates from the direct or 
cascade production of a pair of $Z_1$ boson via the decay mode $Z_1 \rightarrow l_1 l \rightarrow l 
l \gamma_1$, although, the direct production rate is smaller than the cascade production rate. 
The dominant SM backgrounds to the four lepton signal are continuum 
production of $Z^{*}/ \gamma^{*}$ $Z^{*}/ \gamma^{*}$, real ZZ production, heavy flavour 
production, $b\bar{b}b \bar{b}$, $t \bar{t} b \bar{b}$ and  Z production associated with heavy flavour. The cross sections 
of such processes are generally small and by using proper cuts like lepton isolation, Z veto, 
the backgrounds  can be reduced significantly. The four lepton signal is the 
smallest among all of the multilepton states in mUED. It should be remembered that leptons 
coming from the decay of KK electroweak gauge bosons or KK leptons are generally soft. 
In some cases, it is possible to miss one or two of the four leptons. \\

\section{Analysis}
We shall now describe our numerical analysis methods and results. In our scan, we  
vary radius of compactification ($R^{-1}$) from 250 GeV to 1000 GeV although the 
upper limit, consistent with the mUED dark matter bound, is around 1400 GeV.
The reason behind it is that we are trying to discover mUED signals at the LHC in the early
stages of its operation. The luminosity is expected to be small and thus, only signals with 
large cross section will be observed. The total mUED cross section is around 100 fb 
for $R^{-1}$ $\sim$ 1000 GeV. 
We shall also study the discovery reach as a function of cut-off $\Lambda$ of the model. 
Since the value of the cut-off determines the overall splitting 
among KK particles, we expect that mUED search sensitivity will also depend on this parameter. 
Given the value of $R^{-1}$ and the cut-off one can easily calculate the one loop masses of 
the KK particles using the formulae given in the Ref. \cite{mued1}.\\

\subsection{Object Selection}
In our analysis, we have introduce a set of basic selection criteria to identify electrons, muons, jets and missing transverse energy. The object selection is described in brief in the following.

\subsubsection{Electrons/Muons}
The following criteria are used to select a electron/muon candidate:

\begin{itemize}
\item The transverse momentum  $p_T$ of the electron or muon 
must be greater than 10 GeV and pseudorapidity $|\eta|\le2.5$.

\item We require that the leptons be isolated from hadronic activity in the detector.
To select isolated electrons/muons, we demand that the sum of calorimeter energy within a 
distance $\Delta R=0.2$ (where $\Delta R=\sqrt{\Delta \eta^2+\Delta \phi^2}$) around the 
electron/muon candidate must be less than $10$ GeV.

\item We also use lepton jet isolation cut. We remove electrons that are 
found within a distance $0.2 \le\Delta R \le 0.4$ from the jet. 
We also remove all  muon candidates fall within a distance $\Delta R=0.4$ from a jet.
\end{itemize}

\subsubsection{Jets}
Jets are reconstructed with the help of the inbuilt PYTHIA  cluster routine PYCELL. 
PYCELL uses a fixed-size cone algorithm for clustering. 
We have use a cone radius of $\Delta R=0.4$. The following selection criteria are 
introduced to select a jet candidate.

\begin{itemize}
\item The minimum $p_T$ of a jet is taken to be $20$ GeV and we also demand $|\eta_j|<2.5$.
\item Since electrons are likely to be reconstructed also as jets, we remove any jet 
candidate that falls within a distance $\Delta R=0.2$ of a electron candidate.
\end{itemize}  

\subsubsection{Missing transverse energy, $E_T\!\!\!\!\!\!/~~$}
Neutrinos and the $\gamma_1$ escape the detector, leading to
significant missing transverse energy.
The missing transverse energy in an event is calculated using calorimeter cell energy and 
the momentum of the reconstructed muons in the muon spectrometer. 
In our analysis, we have used the very simplified following definition for the missing transverse energy:

\begin{equation}
E_{T}\!\!\!\!\!\!/~~=\sqrt{(\sum p_x)^2+(\sum p_y)^2},
\label{met}
\end{equation}
where, the sum goes over all the isolated electrons, muons, the jets. 

\subsection{Signal distributions}
In this subsection, we briefly study various distributions like missing transverse energy ($E_T\!\!\!\!\!\!/~~$), 
hardness of the jets and leptons, jet multiplicity etc. of the mUED model. We first examine 
missing transverse energy distribution. In the generic SUSY search, one of the most powerful 
discriminator of the signal from the SM backgrounds is the missing transverse energy. We have 
seen in the previous section that mUED 
production cross section at the LHC is dominated by the $n=1$ KK quarks and gluons which decay 
through a chain to the lightest KK particle $\gamma_1$. In the KK parity conserving 
model, $\gamma_1$ escapes the detector and gives rise to missing 
transverse energy. Thus $n=1$ particle production is always associated with $E_T\!\!\!\!\!\!/~~$. 
 However, since we are considering  nearly degenerate KK spectrum, the final state particles will be
soft and thus, it is expected that the $E_T\!\!\!\!\!\!/~~$ distribution will be softer than a typical SUSY $E_T\!\!\!\!\!\!/~~$
distribution. 
\begin{figure}[t]
\hskip 10pt
\epsfig{figure=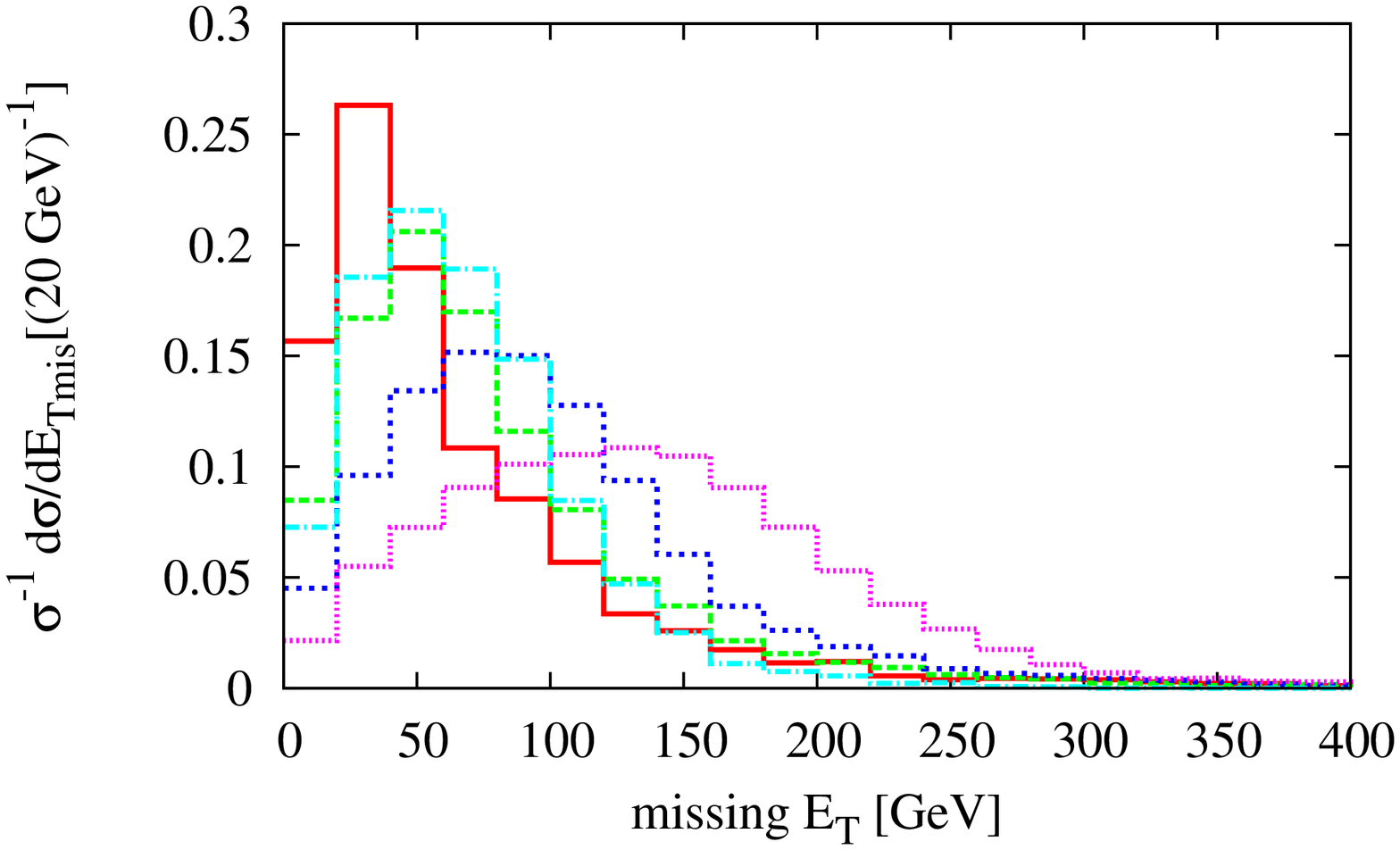,height=10cm,width=10cm,angle=-90}
\hskip -20pt
\epsfig{figure=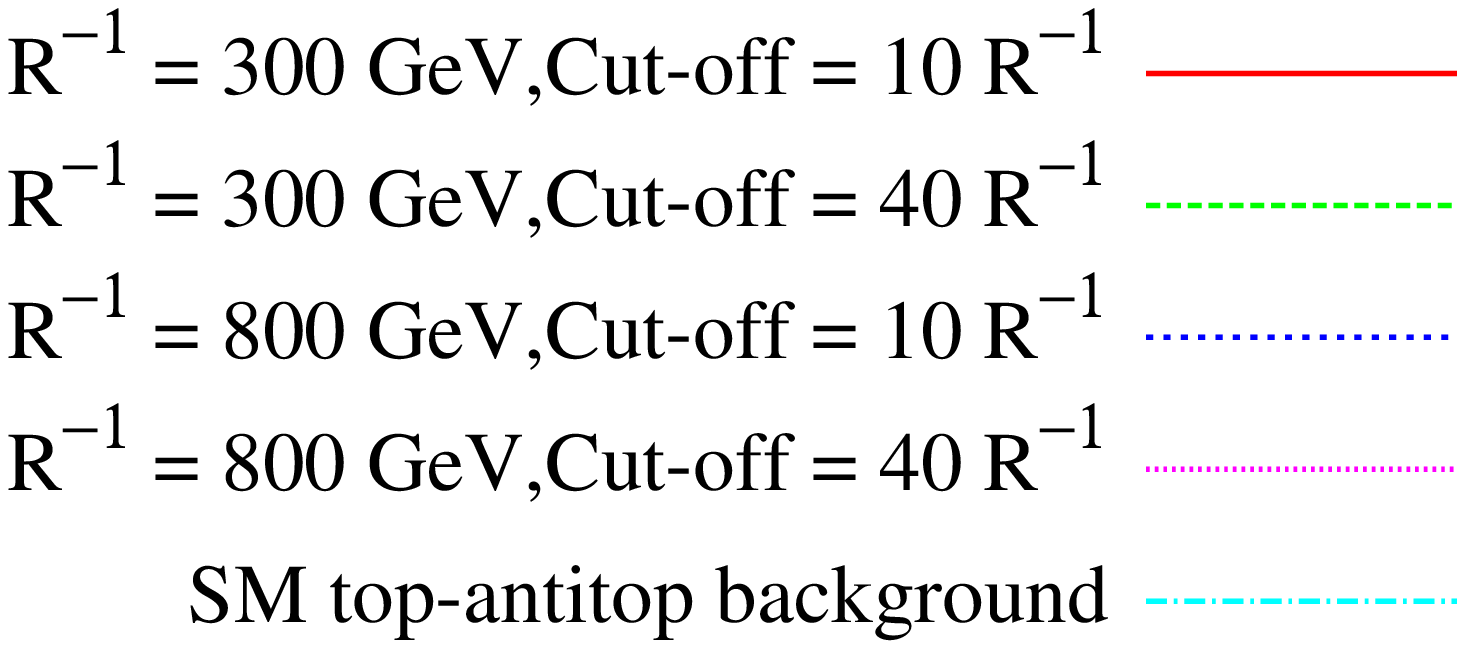,height=8cm,width=8cm,angle=-90}
 \caption{Normalized missing $E_T$ distribution of the events with no. of leptons $\ge 2$  for four signal points: 
($R^{-1},\Lambda R$) $\equiv ~(300,10)$,~$(300,40)$,~$(800,10)$~${\rm and}~(800,40)$ GeV and SM $t \bar{t}$
 background. }
 \label{fig:misspt}
\end{figure}

In Fig.\ref{fig:misspt}, we plot normalized $E_T\!\!\!\!\!\!/$ 
distributions for different values of $R^{-1}$ and $\Lambda R$. We have presented the distributions 
for four different signal points, namely, $(R^{-1},\Lambda R^{-1})~\equiv~(300,10),~(800,10),~(300,40)~
{\rm and}~(800,40)$ GeV. We also show $E_T\!\!\!\!\!\!/~~$ distribution of SM $t\bar{t}$ production. Fig. \ref{fig:misspt} shows 
that, for higher value of $R^{-1}$ and $\Lambda R$, missing $E_T$ increases due to 
the partial removal of the degeneracy of the mUED mass spectrum. 
For $R^{-1}$=300 GeV and $\Lambda R$ =40, the $E_T\!\!\!\!\!\!/~~$ peaks at around 50 GeV,
and it peaks at 130 GeV for $R^{-1}$ =800 GeV and $\Lambda R$=40. In the Standard Model, 
only sources of missing transverse energy are standard model neutrinos. 
Thus, the major SM backgrounds which offer missing transverse energy are 
$W$, $Z$ and $t$ production. Various other effects like mismeasurement of jet 
energy, cracks in the detector can also generate sizeable amount of missing transverse energy.  
 In order to estimate the missing energy originated from such effects, one has to understand the detector 
properly and it will take a good amount of time for detector calibration. In the early stage of LHC 
run, missing energy may not be measured accurately. It is a therefore reasonable to study the reach 
of LHC without using $E_T\!\!\!\!\!\!/~~$ information. The missing $E_T$ distribution of mUED 
is not much different from SM and thus hard cut on the missing  $E_T\!\!\!\!\!\!/~~$ will remove majority of the 
signal events. In this paper, we have demonstrated two different possibilities , e.g., (a). reach of LHC 
with soft $E_T\!\!\!\!\!\!/~~$ cut, (b). LHC reach 
without using missing $E_T$ cut on the signals as well as backgrounds. For better signal significance, one can optimize the cut on the 
missing $E_T$ and hardness of the leptons and jets depending on the masses of the KK particles however, no such attempt is made in our analysis.

\begin{figure}[h]
\epsfig{figure=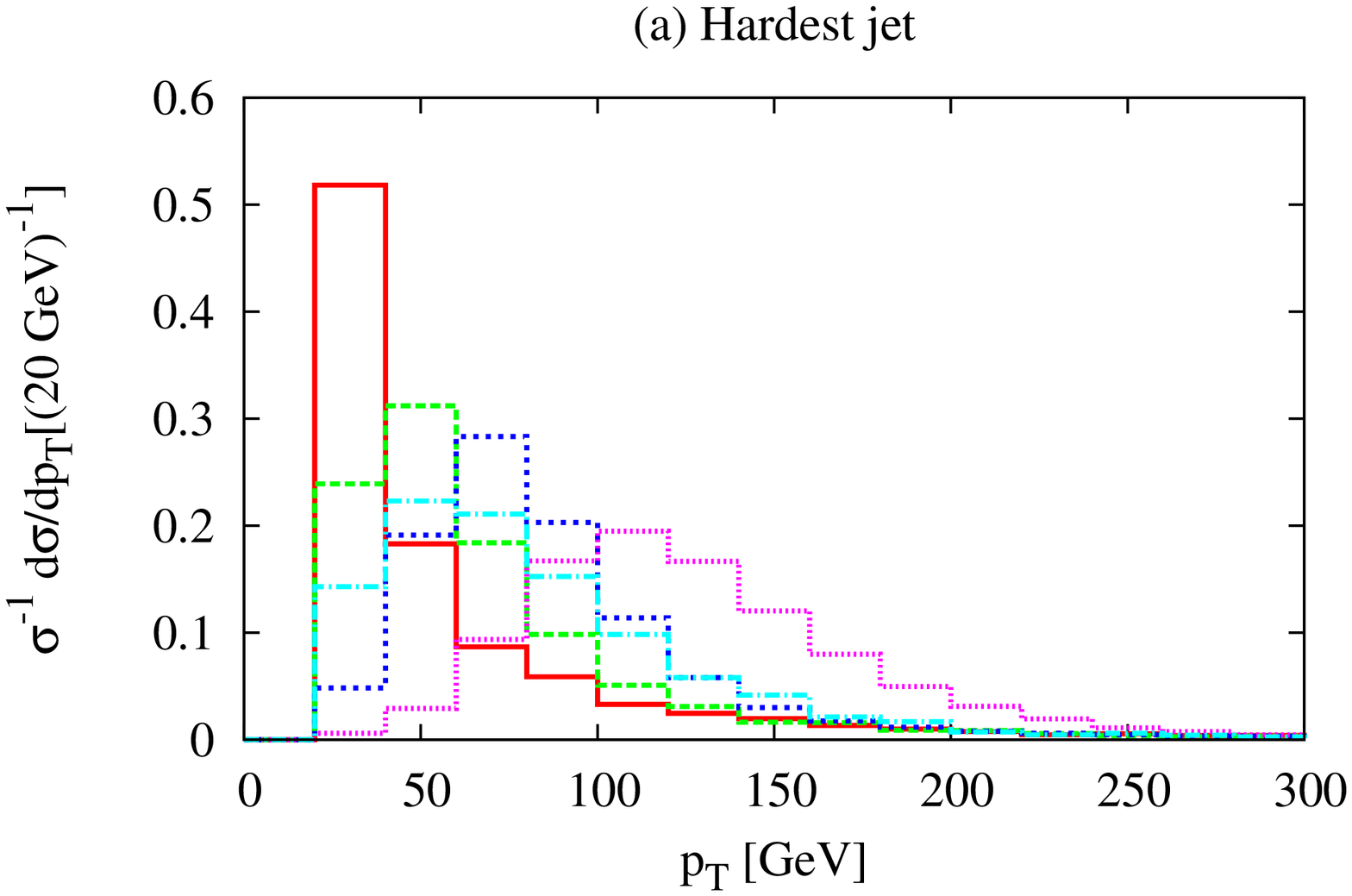,height=5.5cm,width=5.3cm,angle=-90}
\epsfig{figure=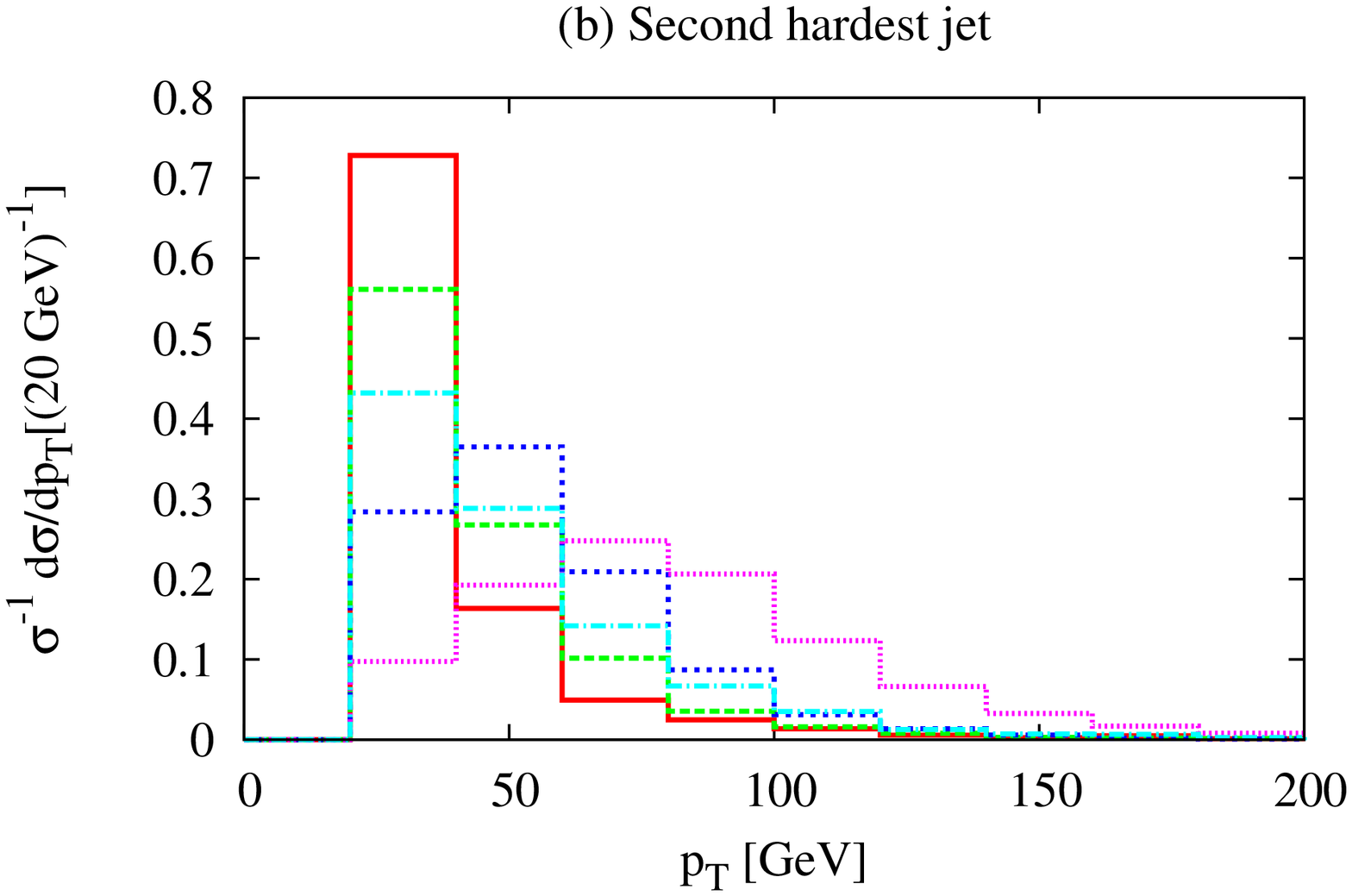,height=5.5cm,width=5.3cm,angle=-90}
\epsfig{figure=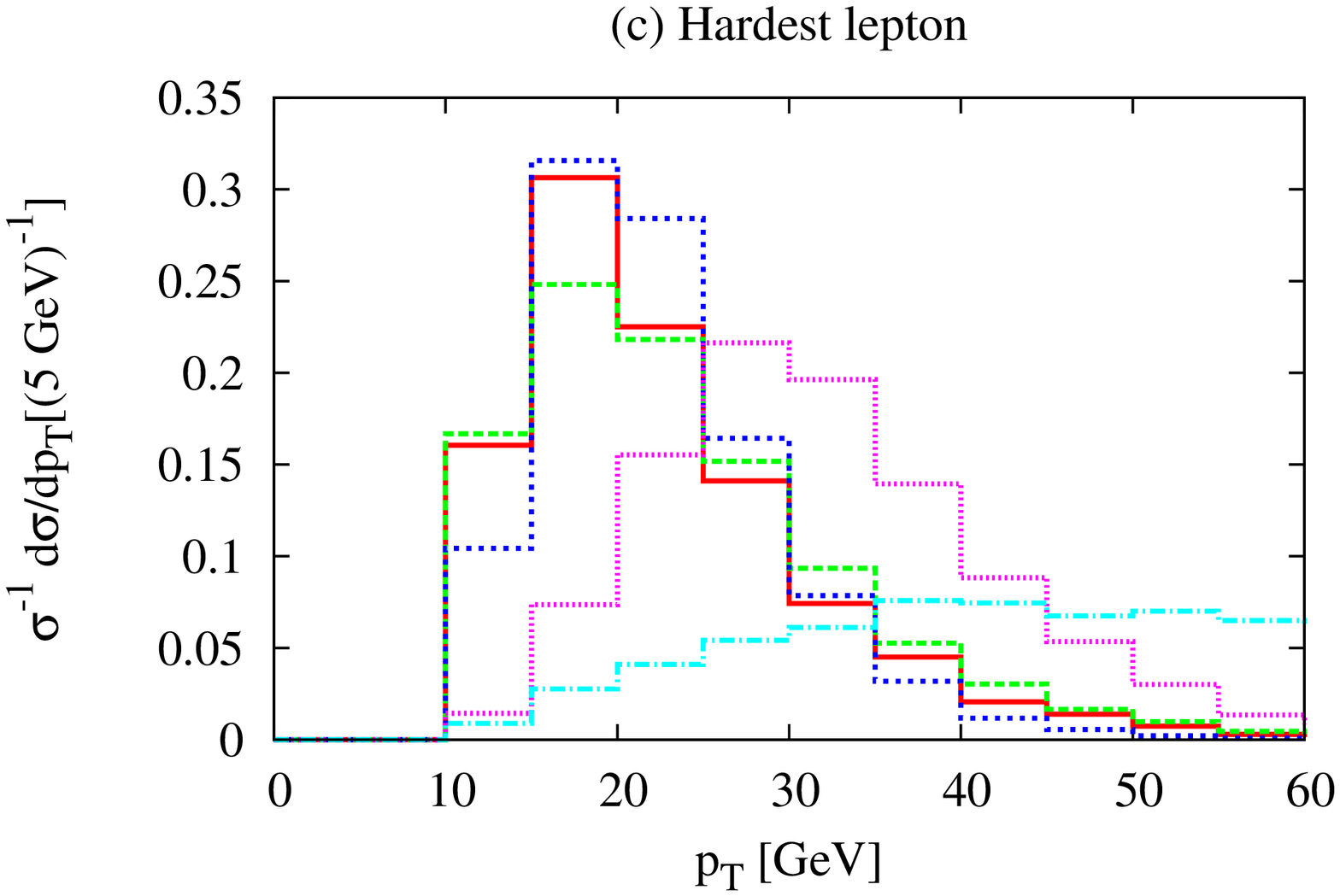,height=5.5cm,width=5.3cm,angle=-90}
\epsfig{figure=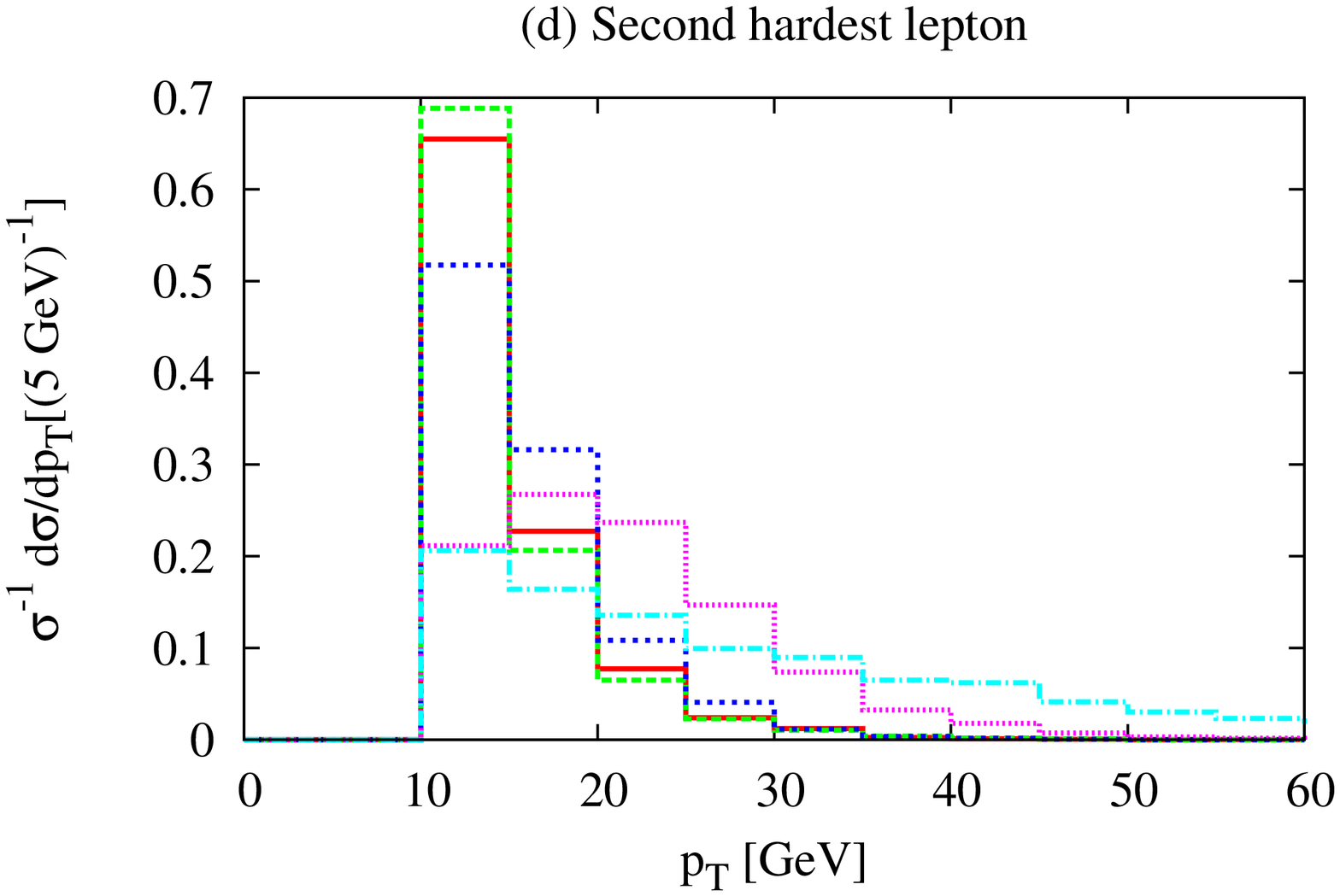,height=5.5cm,width=5.3cm,angle=-90}
\hspace{2cm}
\epsfig{figure=box.ps,height=5.cm,width=5.cm,angle=-90}
\caption{Normalized transverse momentum distributions of the events: (a) the hardest jet, 
(b) the second hardest jet, (c) the hardest lepton and (d) the second hardest 
lepton. Each panel shows distributions for four signal points: 
($R^{-1},\Lambda R$) $\equiv ~(300,10)$,~$(300,40)$,~$(800,10)$~${\rm and}~(800,40)$ GeV and SM $t \bar{t}$ 
background.}
 \label{fig:pTdist}
\end{figure}

We next show the $p_T$ distributions of leptons and jets coming from the decay of KK particles.
Since we are concentrating only on 2, 3 and 4-lepton events, we have used 
all the  events with number of lepton $\ge$ 2 to numerically evaluate those distributions. 
In the Fig. \ref{fig:pTdist}, we have presented the normalized transverse momentum 
distributions of (a) the hardest 
jet, (b) the second hardest jet, (c) the hardest lepton and (d) the second hardest lepton. 
Fig. \ref{fig:pTdist} clearly suggests that mUED model will give rise to very 
soft leptons and jets at the LHC. This is a consequence of nearly degenerate mUED mass spectrum. 
As we increase the value of $R^{-1}$ or $\Lambda$, the mass splitting between different $n=1$ particles 
increases. Therefore, we will have relatively harder leptons and jets for large value of $R^{-1}$ or 
$\Lambda$. The $p_T$ distribution of leptons show that hardest lepton $p_T$ does not exceed 60 GeV 
and we can use a upper cut of 50 GeV on the hardest lepton's transverse momentum to reject Standard 
Model backgrounds.

\noindent

\begin{figure}[h]
\epsfig{figure=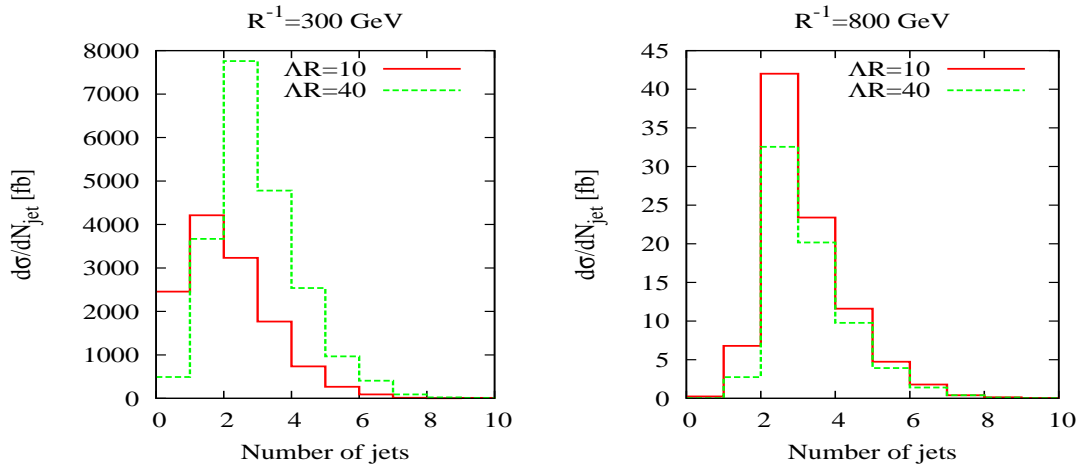,height=15cm,width=8cm,angle=-90}
\caption{Jet multiplicity distributions for four different signal points. 
Left panel: ($R^{-1},\Lambda R$) $\equiv ~(300,10)$,~$(300,40)$ GeV and 
right panel: ($R^{-1},\Lambda R$) $\equiv ~(800,10)$,~$(800,40)$ GeV.}
\label{fig:njet}
\end{figure}

In the Fig. \ref{fig:njet}, we have presented the jet multiplicity distribution of signal events. 
Since we have considered only QCD production channels (namely, $g_1g_1,~g_1Q_1,~ {\rm and}~ Q_1Q_1$), 
all the events are accompanied by one or more jets. 
In fact, more than 79 (96)\% of the multilepton events for $R^{-1}=300~(800)$ GeV 
 and $\Lambda R=40$ are accompanied by two or more jets. 
We can exploit this feature of the signal to reduce the contributions from the SM background 
process (like Single $W,~Z$-boson, $WW$, $ZZ$, $WZ$ etc. productions) which may have high lepton 
multiplicity but low jet multiplicity.  
\begin{figure}[h]
\epsfig{figure=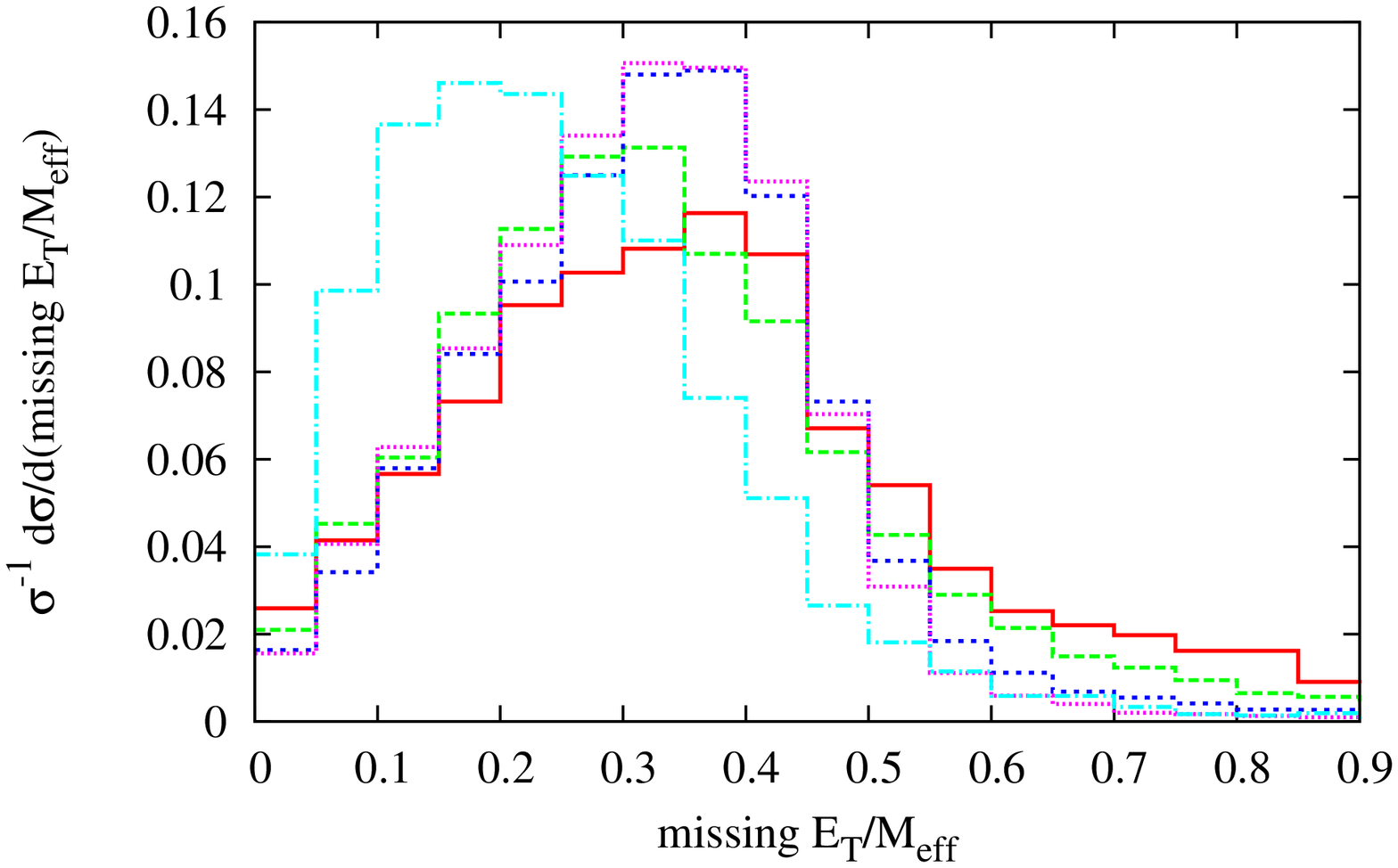,height=10cm,width=10cm,angle=-90}
\hskip -20pt
\epsfig{figure=box.ps,height=7cm,width=7cm,angle=-90}
\caption{Normalized $\frac{E_T\!\!\!\!\!\!\!/}{M_{eff}}$ distributions for four different signal points and SM $t \bar{t}$ background.}
\label{fig:met_meff}
\end{figure}

In the Fig. \ref{fig:met_meff}, we present the normalized 
$\frac{E_T\!\!\!\!\!\!\!/}{M_{eff}}$ distributions for four different signal points. A lower bound on the 
$\frac{E_T\!\!\!\!\!\!\!/}{M_{eff}}$ could be very efficient to reduce 
background contributions where $E_T\!\!\!\!\!\!\!/~~$ arises from the mis-measurement 
of visible momenta. 

\begin{table}[htbp]
\begin{center}
\centering
\begin{tabular}{||l|c|c||}
\hline\hline
Process                 & Generator     &    Cross section in [pb]  \\
   &      & $\sqrt s=7$ TeV \\    
\hline\hline
$t \bar{t}$: $t \bar t~+~0,~1~{\rm and} ~2$ jets           & PYTHIA+ALPGEN & 113.01 \\
$WW$: $W^{\pm}(\to l\nu)+W^{\pm}(\to l\nu)~+~0,~1~{\rm and}~2$ jets & PYTHIA+ALPGEN & 4.23 \\
$ZZ$: $Z(\to all)+Z(\to all)~+~0,~1~{\rm and}~2$ jets & PYTHIA+ALPGEN & 6.75 \\
$WZ$: $W^{\pm}(\to all)+Z(\to all)~+~0,~1~{\rm and}~2$ jets & PYTHIA+ALPGEN & 15.09 \\
$W+b\bar b$: $W^{\pm}(\to all)+b \bar b$ & PYTHIA+ALPGEN & 3.22\\
$W+t\bar t$: $W^{\pm}(\to all)+t \bar t$ & PYTHIA+ALPGEN & 1.39$\times 10^{-1}$ \\
$Z+b\bar b$: $Z(\to l\bar l)+b \bar b$ & PYTHIA+ALPGEN & 1.6$\times 10^{-1}$\\
$Z+t\bar t$: $Z(\to l\bar l)+t \bar t$ & PYTHIA+ALPGEN & 4.37$\times 10^{-3}$ \\
$W/Z/\gamma$:  & PYTHIA                & 4.23$\times 10^{5}$\\
$W+$jets: $W^{\pm}(\to l\nu)+1$ jet & PYTHIA+ALPGEN & 1.00$\times 10^{3}$ \\
$Z+$jets: $Z(\to l\bar l)+1$ jet & PYTHIA+ALPGEN & 30.9 \\
$t \bar{t} t \bar{t}$  &   PYTHIA+CALcHEP              &    less than 0.001         \\ 
$b \bar{b} b \bar{b}$  &   PYTHIA+CALcHEP             &     14             \\ 
$t \bar{t} b \bar{b}$  &   PYTHIA+CALcHEP           &      0.14                \\ 
\hline\hline
\end{tabular}
\caption{ List of Standard Model backgrounds used in our analysis. The cross 
sections are evaluated with $\sqrt{s}$=7 TeV, using PYTHIA,ALPGEN and CALcHEP.  }
\label{smcross}
\end{center}
\end{table}

As already explained, the Standard Model background processes most relevant to mUED searches in multilepton 
mode are $t \bar{t}$, single top, $W +jets$, $Z+jets$ and diboson 
production. We use PYTHIA to generate signal events and for SM backgrounds ALPGEN \cite{ALPGEN} and CALcHEP \cite{Pukhov} is used. 
The parton level unweighted events from ALPGEN or CalCHEP has been interfaced with PYTHIA. We use the 
CTEQ5M parton distribution function in 
our analysis. The list of the SM background process considered and their cross sections are presented in Table \ref{smcross}. We have used  
inbuilt initial and final state radiations (ISR/FSR), fragmentation and hadronization routines in the 
PYTHIA code and we do not consider the effect of multiple interactions. \\

\subsection{Event Selection}
After discussing about the signal and background characteristics, we are now equipped enough to 
introduce a set event selection criteria which will enhance the signal to background ratio. 
We have applied channel independent as well as channel dependent cuts. We 
have used  b tagging with efficiency $\epsilon_b=0.5$ to minimize SM top quark backgrounds. 
The event selection criteria for different signal 
topologies under consideration are summarized in the following. 

\subsubsection{Same-sign di-lepton (SSD) events}

\begin{itemize}
\item We demand exactly one pair of leptons with same charge and $p_T>10$ GeV. 
We reject events with a third lepton with $p_T>10$ GeV.

\item As discussed in the previous section, the softness of the signal leptons can also be 
utilized to enhance signal to background ratio. We demand $p_T^{l_1}<50$ GeV and 
$p_T^{l_2}<30$ GeV where $l_1~{\rm and}~l_2$ are the hardest and second hardest lepton respectively.

\item We require at least two jets with $p_T>40$ GeV. This cut is very efficient in 
reducing the contribution from SM background which may have high lepton multiplicity but 
low jet multiplicity.

\item Since the SM events are more likely to have lower $E_T\!\!\!\!\!\!/~~$, a higher 
bound on the $E_T\!\!\!\!\!\!/~~$ would be very effective against QCD di-jet and 
$Z+$jets events. However, as discussed earlier, due to 
the close degeneracy of the mUED mass spectrum, we will not have 
large $E_T\!\!\!\!\!\!/~~$ for signal events. Therefore, we 
demand $E_T\!\!\!\!\!\!/~~$ to be greater than $50$ GeV in all events. 

\item We require missing $E_T >  f \times M_{eff}$ where f = 0.2. 

\item As mentioned earlier, the most dominant SM background process is the 
$t \bar t$ production followed by the semi-leptonic decay of one b-quark, 
leptonic decay of one $W$-boson and hadronic decay of the other $W$-boson. 
$t \bar t$ contribution can be suppressed by rejecting events with 
tagged b jets. Therefore, we use b jet veto to minimize the SM top backgrounds.

\end{itemize}

\subsubsection{Opposite-sign di-lepton (OSD) events}

\begin{itemize}

\item We require the presence of exactly two leptons with opposite charge and $p_T > 10$
GeV. 
\item We impose an upper bound of 50 and 30 GeV on the transverse momentum of the hardest and 
second hardest lepton respectively.
\item  We require the missing $E_T$ to be larger than 50 GeV.
\item At least 2 jets with $p_T$ $>$ 40 GeV.
\item Missing $E_T >  f \times M_{eff}$ where f = 0.2.
\item  We also apply b jet veto to minimize SM top backgrounds. 
\item  We vetoed events exhibiting the same flavour opposite charge lepton pair 
with an invariant mass smaller than 10 GeV and also invariant mass between 80 to 100 GeV.
This cut removes both Z boson and QED contribution.   
\end{itemize}

\subsubsection{Tri-lepton events}

\begin{itemize}

\item We require the presence of exactly three isolated leptons with $p_T$ greater than 10 GeV. 
\item We demand $p_T^{l_1}<50$ GeV and $p_T^{l_2}<30$ GeV. 
\item  Missing transverse energy $E_T\!\!\!\!\!\!/~~>$ 50 GeV. 
\item  We require atleast 2 jets with $p_T>$ 40 GeV. 
\item  The ratio of missing $E_T$ and effective mass must be greater than 0.2. 
\item We put the same cut on the invariant mass on the opposite sign same flavour (ossf) leptons as described 
above, i.e., $|M_{ossf}-M_Z| >$ 10 GeV, and $M_{ossf}> 10$ GeV.    
\item We also apply b jet veto. 
\end{itemize}

\subsubsection{Four-lepton events}

\begin{itemize}
\item We require 4 leptons with $p_T$ greater than 10 GeV. We also require the hardest and 
second hardest lepton transverse momentum to be less than 50 and 30 GeV respectively.
\item Missing $E_T>$ 50 GeV.
\item Atleast 2 jets with $p_T>$ 40 GeV.
\item We also apply the same invariant mass cut on the opposite sign same flavour leptons 
and b jet veto as described above.
\end{itemize}

\begin{figure}[t]
\epsfig{figure=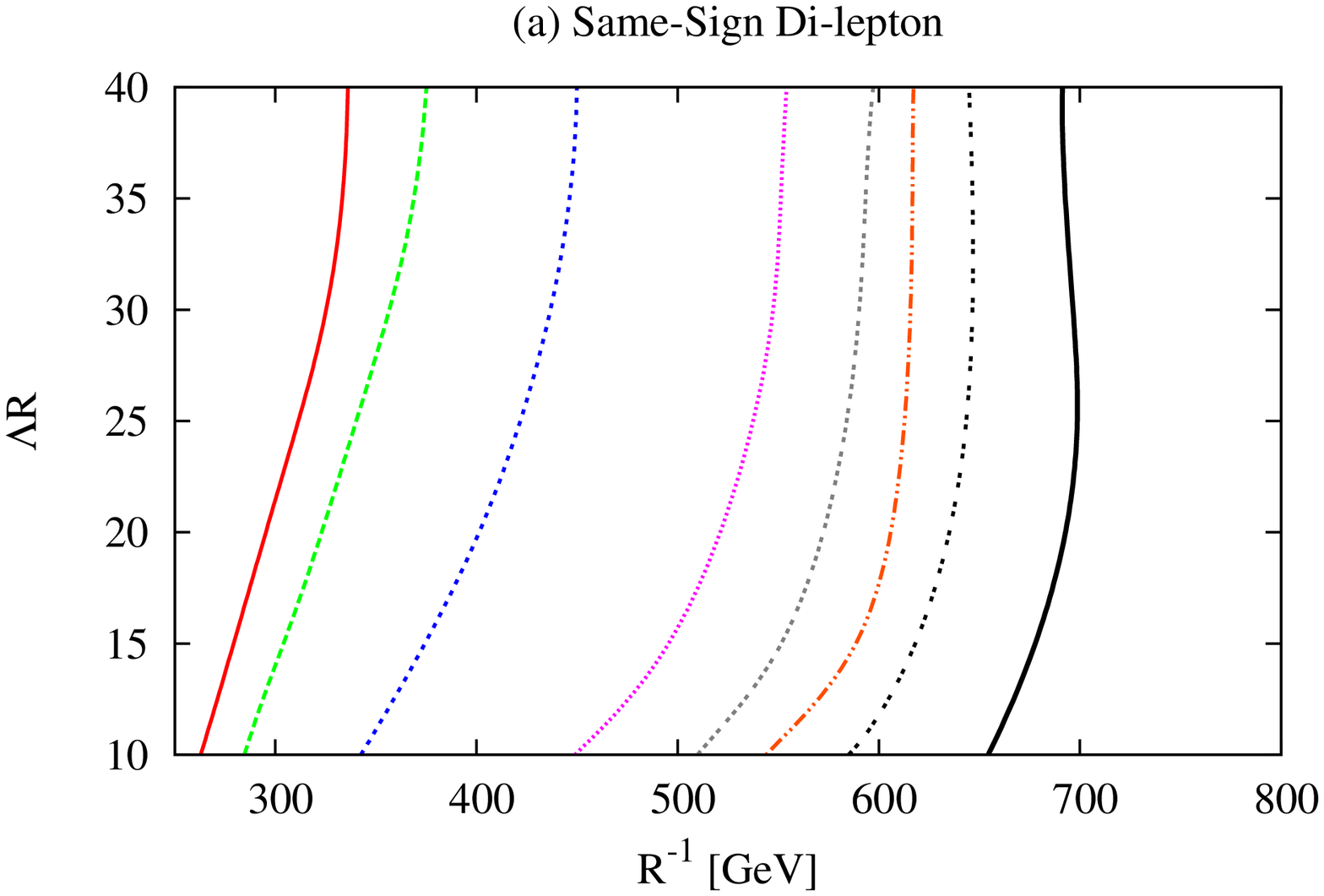,height=8.8cm,width=6.5cm,angle=-90}
\epsfig{figure=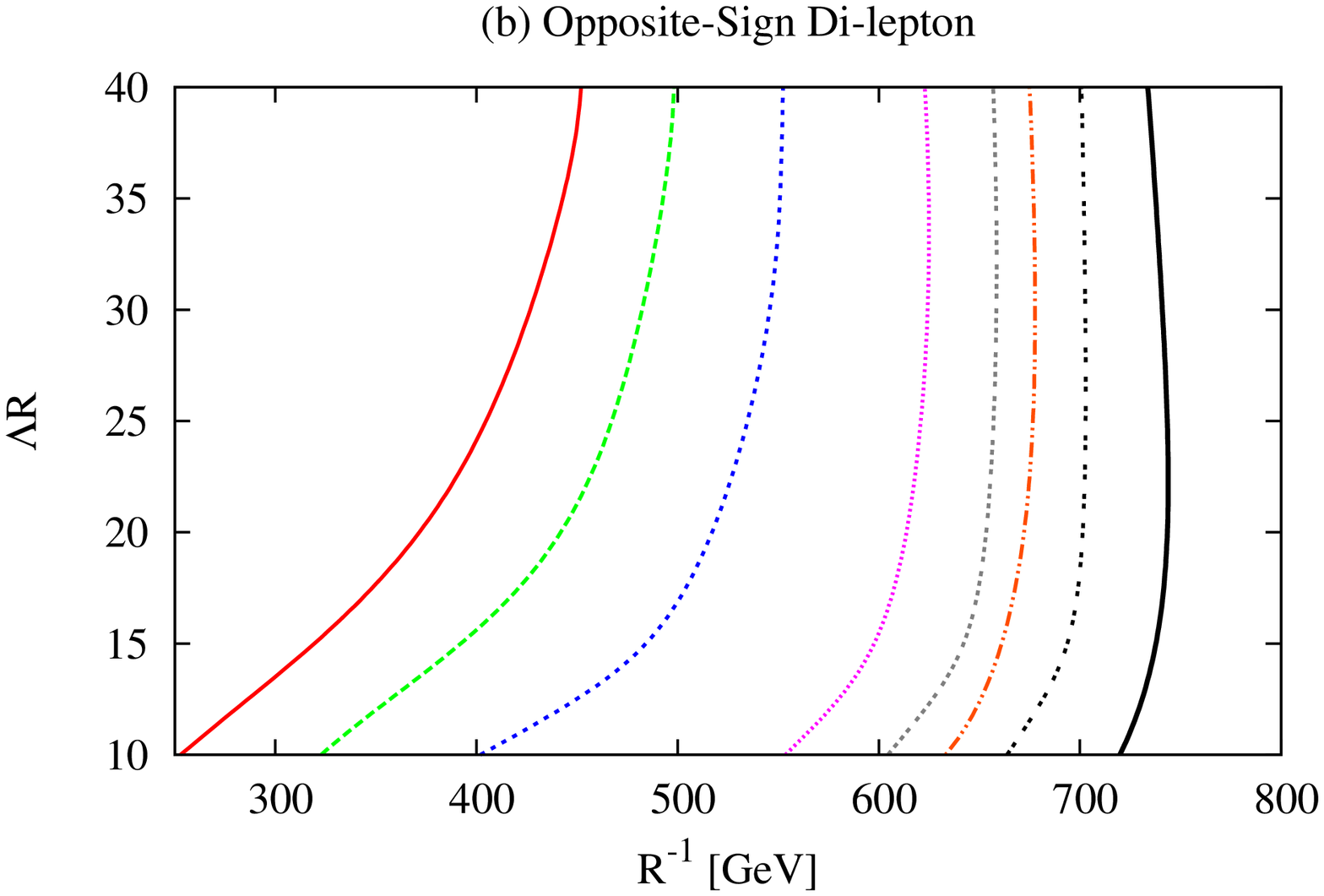,height=8.8cm,width=6.5cm,angle=-90}
\epsfig{figure=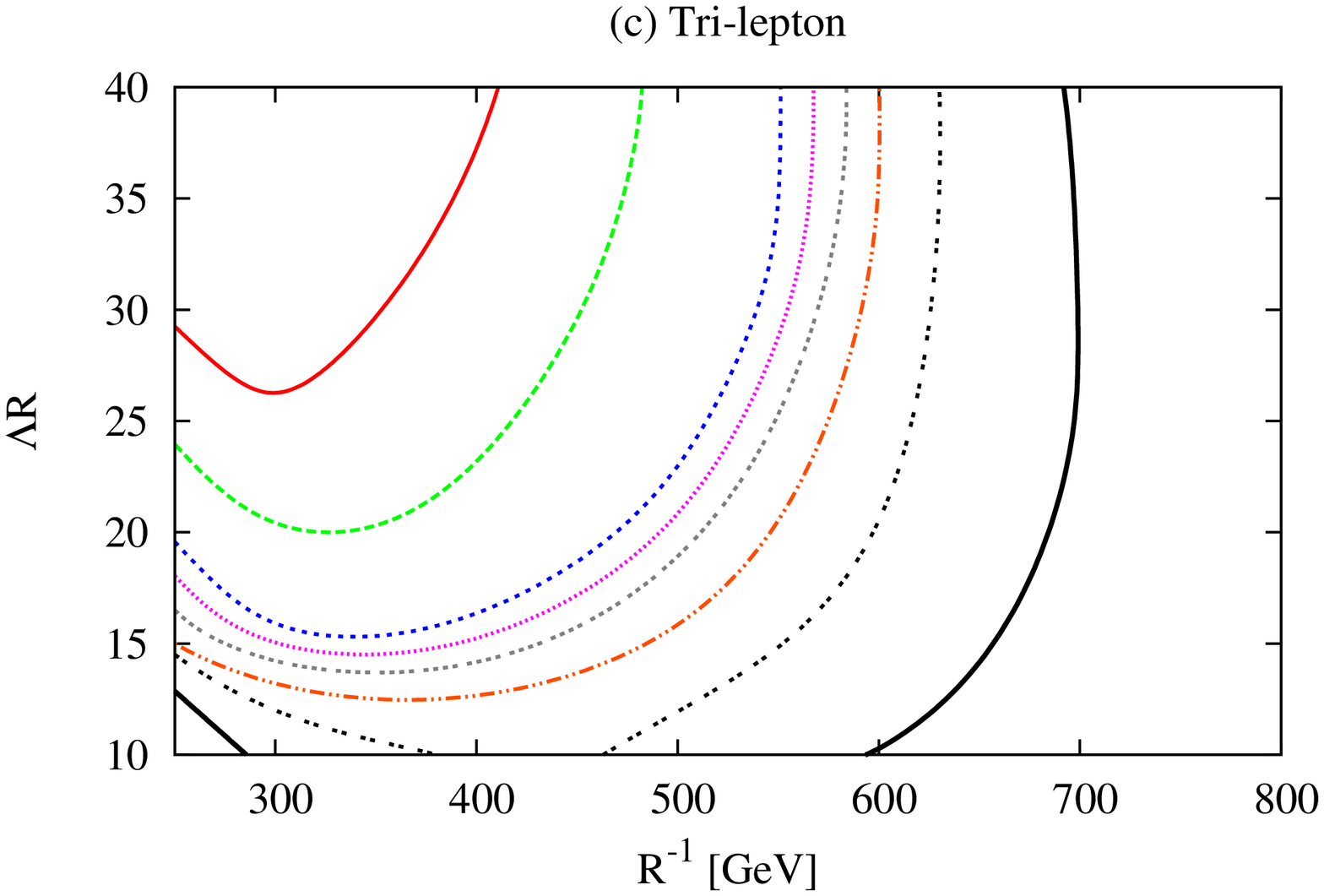,height=8.8cm,width=6.5cm,angle=-90}
\hspace{2cm}
\epsfig{figure=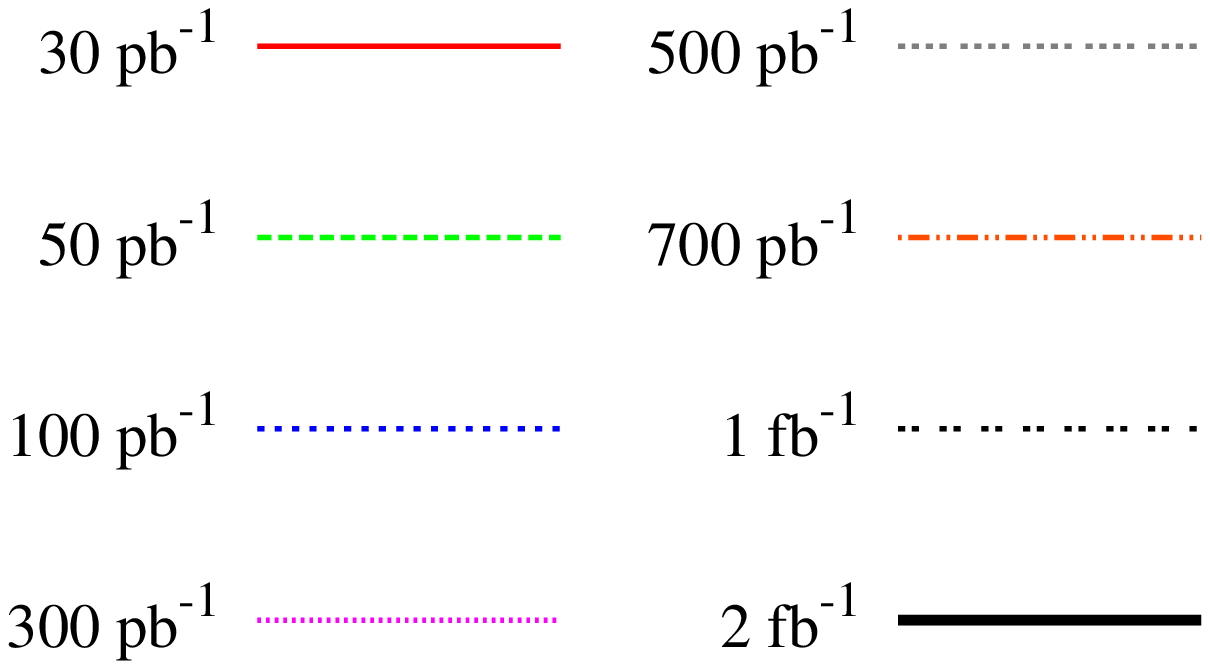,height=8.8cm,width=6.5cm,angle=-90}
\caption{Discovery reach of mUED at the LHC with $\sqrt s=7$ TeV. (a). Same-sign di-lepton, 
(b). opposite-sign di-lepton and (c) tri-lepton channel. We have used $E_T\!\!\!\!\!\!/~~$ to 
suppress the SM backgrounds. Eight different values of integrated luminosities are assumed. 
Each line corresponds to $5\sigma$ contour in the $R^{-1}-\Lambda R$ plane.}
 \label{fig:reach_with}
\end{figure}

\subsection{Discovery reach}
\noindent In this section, we will concentrate on the discovery potential of mUED model at the LHC with centre of mass energy 7 TeV where the maximum integrated luminosity is likely to be of the order of 1 to 2 $fb^{-1}$. 
The masses of the KK particles scale linearly with $R^{-1}$ and hence, total KK cross section
decreases with it. We choose $R^{-1}$ as one of the scan parameter. We have chosen $\Lambda R$ as the 
second scan parameter because it determines the splitting between two KK states. 
The determination of $E_T\!\!\!\!\!\!/~~$ crucially depends on the precise knowledge of the detector response. 
Since fake $E_T\!\!\!\!\!\!/~~$ grows with the total scalar energy in hadron collider events, the knowledge about the detector performance at $\sqrt s \sim 0.95-2.36$ TeV may not be useful. Therefore, at the early stage of LHC with $\sqrt s=7$ TeV, precise information about the $E_T\!\!\!\!\!\!/~~$ may not be available. Keeping this in mind, we have divided our analysis into two different categories. For category-1, we assume that there will be precise information about $E_T\!\!\!\!\!\!/~~$. We use this information to suppress the SM backgrounds. For category-2, we do not use any information about $E_T\!\!\!\!\!\!/~~$.

First, we show the discovery reach in the multileptonic channel using the $E_T\!\!\!\!\!\!/~~$ i.e., category-1. 
We define the signal to be observable for a integrated luminosity ${\cal L}$ if,

\begin{itemize}
\item
\begin{equation}
\frac{N_{S}}{\sqrt{N_B+N_S}} \ge 5 ~~~~~~ {\rm for}~~~~~ 0< N_B \le 5 N_S,
\end{equation}
where, $N_{S(B)}=\sigma_{S(B)} {\cal L}$, is the number of signal (background) events for an integrated 
luminosity ${\cal L}$.
\item For zero number of background event, the signal is observable if there are at 
least five signal events. 
\item In order to establish the discovery of a small signal (which could be 
statistically significant i.e. $N_S/\sqrt{N_B}\ge 5$) on top of a large background, we need to know the 
background with exquisite precision. However, such precise determination of the SM background is beyond 
the scope of this present article. Therefore, we impose the requirement $N_B\le 5 N_S$ to avoid such 
possibilities.

\end{itemize}

In Fig. \ref{fig:reach_with}, we have presented the mUED discovery potential of 
the LHC in the multilepton plus $E_T\!\!\!\!\!\!/~~$ channels namely, (a). same-sign di-lepton, 
(b). opposite-sign di-lepton and (c) tri-lepton channel. We have assumed eight different 
values of integrated luminosity ranging over $30~{\rm pb}^{-1}$ to $2~{\rm fb}^{-1}$. 
Figure shows the 5 $\sigma$ discovery contour in the $R^{-1}$-$\Lambda R$ plane and the lines refer to the 
different integrated luminosities.

\begin{itemize}

\item Fig. \ref{fig:reach_with} clearly suggests that for low integrated luminosity OSD and tri-lepton 
channels are most promising. As for example, with $100~{\rm pb}^{-1}$ integrated luminosity, 
one can probe mUED parameter space upto $R^{-1}=550 $ GeV for $\Lambda R=30$ in the OSD or 
tri-lepton channel. Whereas, the corresponding discovery reach in SSD channel is $R^{-1}=440$ 
GeV. 

\item Fig. \ref{fig:reach_with} shows that discovery potential in these channels strongly depend on 
the $\Lambda R$ of the model for low value of $R^{-1}$ and it is almost independent 
of $\Lambda R$ in case of high value of $R^{-1}$. 
It is also clear from the Fig. \ref{fig:reach_with} that probing lower values of the cut-off scale 
($\Lambda$) will be difficult i.e. we require higher integrated luminosity to probe low 
$\Lambda R$ region. This is a consequence of the fact that lower value cut-off scale will 
corresponds to more degenerate mUED mass spectrum. If the mass-splitting between different 
$n=1$ KK states are small, the final state leptons and jets (arising from the cascade decay 
of $n=1$ quarks and gluons) will be very soft and fall out side of the detector acceptance quite 
often. Therefore, for small $\Lambda R$, the signal cross-section decreases and we require 
higher integrated luminosity to probe the region. This feature can be observed in all the 
three channels, however, most prominent in the tri-lepton channel. Fig. \ref{fig:reach_with}c shows
 that in the tri-lepton channel, one can not probe $\Lambda R<20$ region with $50~{\rm pb}^{-1}$ integrated 
luminosity. Since the mass splitting increases with in increasing values of $R^{-1}$, 
the sensitivity of $\Lambda R$ decreases for higher values of $R^{-1}$. 

\item For  high integrated luminosity, OSD is the most promising channel and it can probe $R^{-1}$ $\sim$
750 (700) GeV at an integrated luminosity of $2~(1)~{\rm fb}^{-1}$. This reach is almost independent of the $\Lambda R$. The SSD and tri-lepton channels have less background, but the 
signal cross sections are also small. The OSD channel, background being an order of magnitude larger, is on the 
other hand more useful. Hence, the OSD channel results may actually be the most important for the 
early searches.

\end{itemize}

\begin{figure}[h]
\epsfig{figure=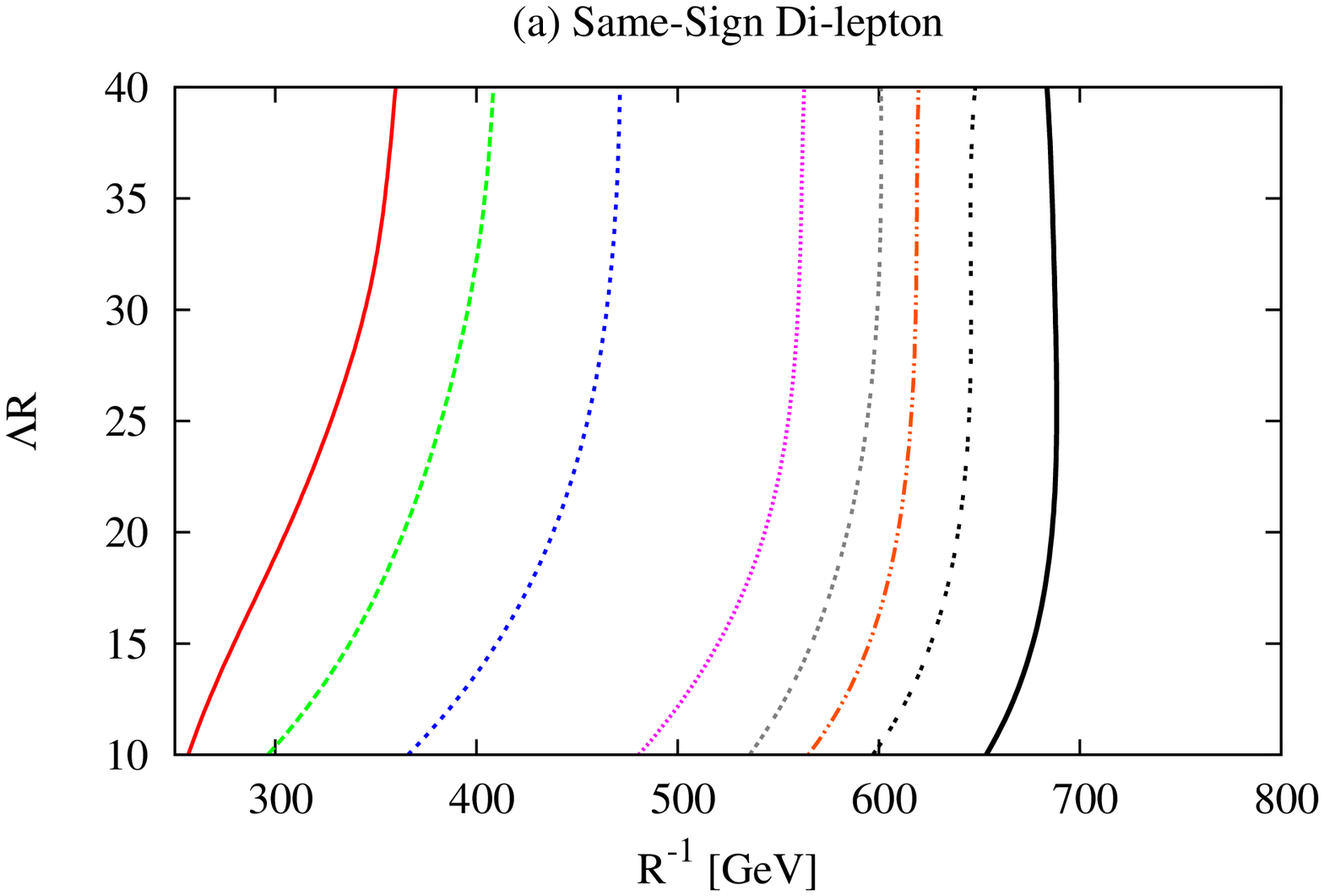,height=8.8cm,width=6.5cm,angle=-90}
\epsfig{figure=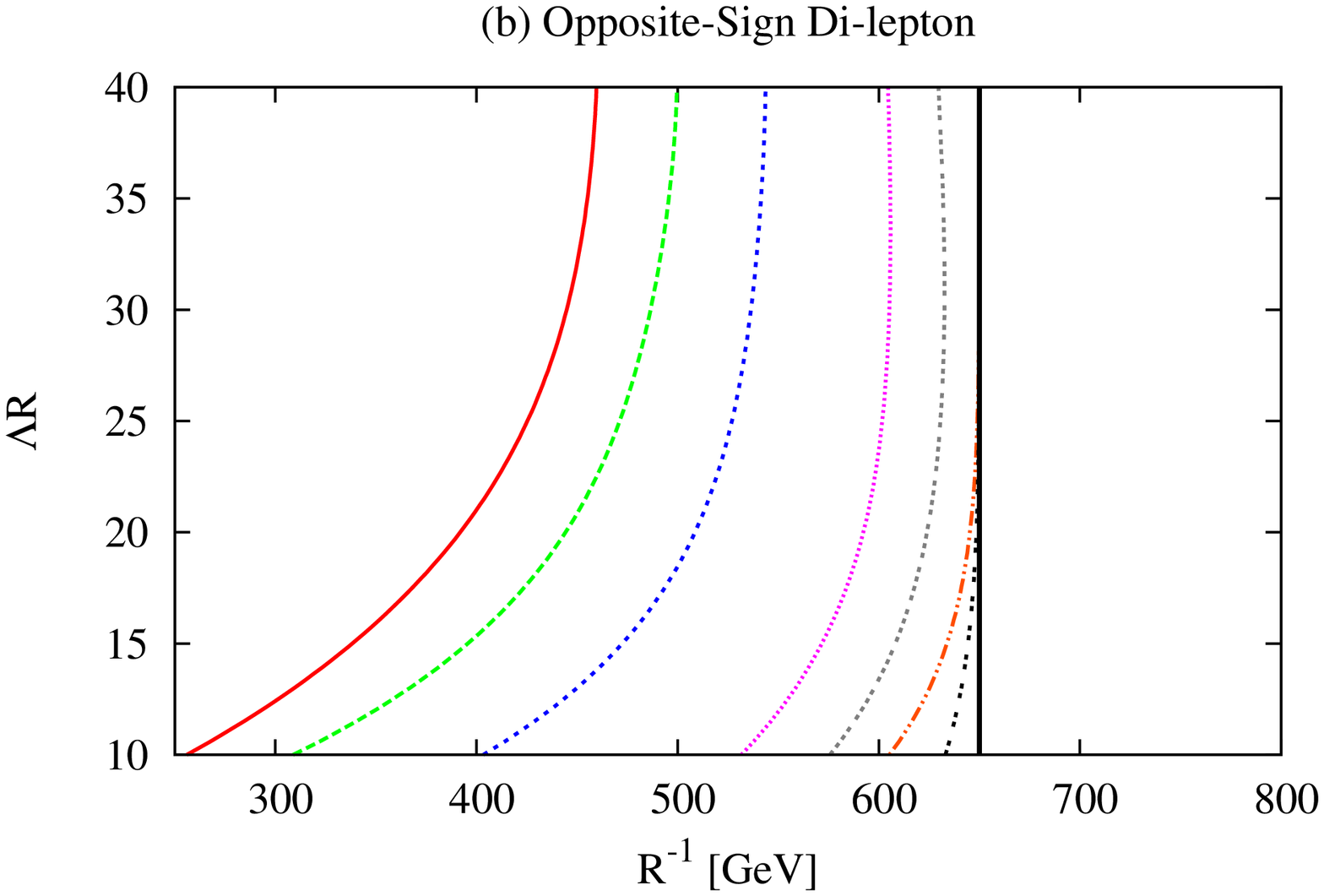,height=8.8cm,width=6.5cm,angle=-90}
\epsfig{figure=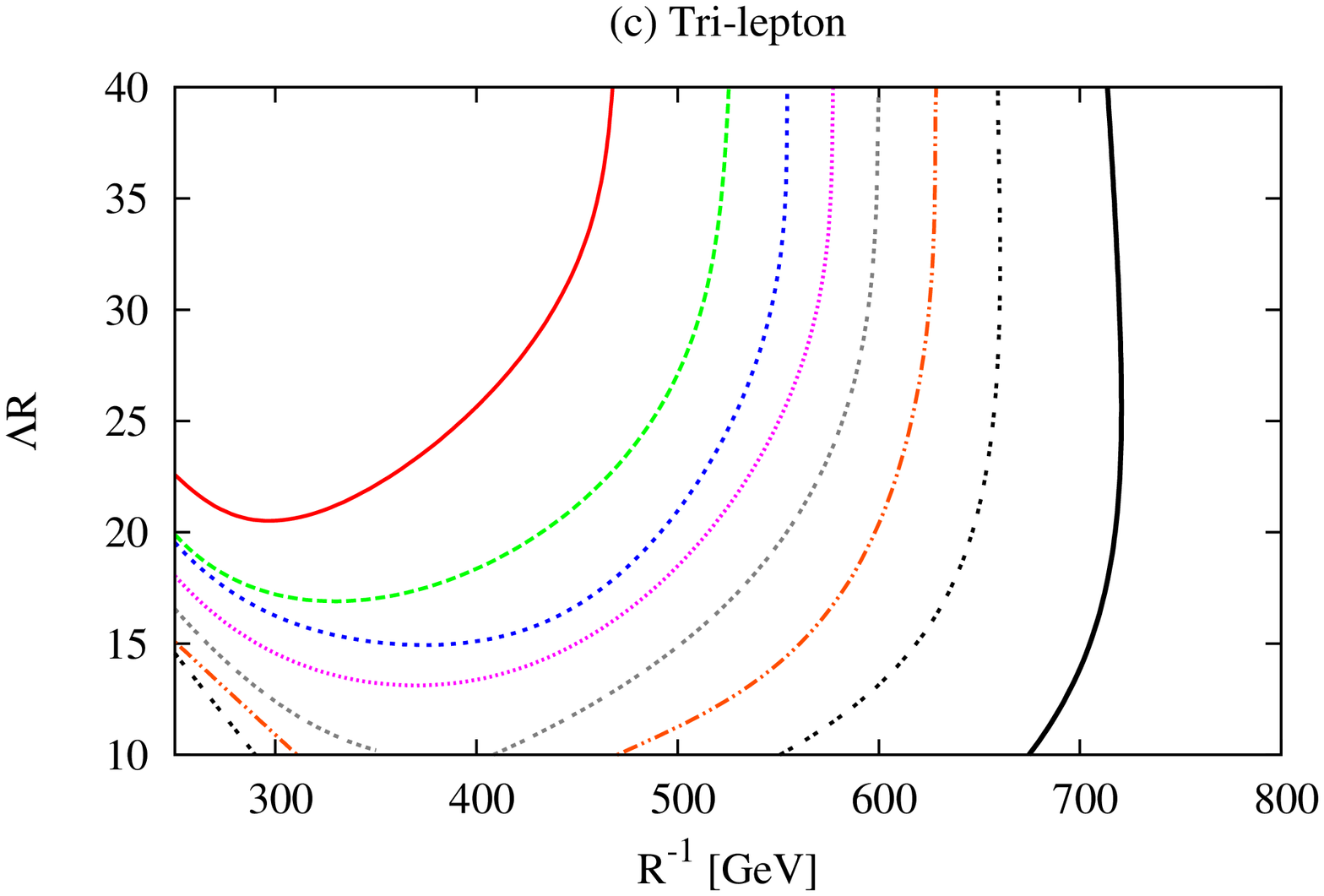,height=8.8cm,width=6.5cm,angle=-90}
\hspace{2cm}
\epsfig{figure=reach_plot/label.ps,height=8.8cm,width=6.5cm,angle=-90}
\caption{Discovery reach of mUED at the LHC with $\sqrt s=7$ TeV. (a). Same-sign di-lepton, 
(b). opposite-sign di-lepton and (c) tri-lepton channel. We do not use any information about the $E_T\!\!\!\!\!\!/~~$. 
Eight different values of integrated luminosities are assumed. 
Each line corresponds to $5\sigma$ contour in the $R^{-1}-\Lambda R$ plane.}
 \label{fig:reach_without}
\end{figure}

As stated earlier, the understanding of the nature of the $E_T\!\!\!\!\!\!/~$ will be crucial in the early 
search for new physics at the LHC. At the early stage of LHC, estimation and the rejection 
of the background from physics processes  with no real intrinsic transverse missing energy is  
more challenging because the measured $E_T\!\!\!\!\!\!/~~$ would not reflect the one originating from the 
non-interacting particles. For this reason, we have carried out the same analysis without 
considering the $E_T\!\!\!\!\!\!/~~$ cut. In the Fig. \ref{fig:reach_without}, we present the discovery 
reach of mUED model without using
$E_T\!\!\!\!\!\!/~~$\footnote{In this part of our analysis, we do not use $E_T\!\!\!\!\!\!/~~>50$ GeV 
and $E_T\!\!\!\!\!\!/~~>0.2\times M_{eff}$ cuts.}. The $E_T\!\!\!\!\!\!/~~$ distribution of 
this model is soft (see Fig. \ref{fig:misspt}) and therefore, it is expected that inclusion 
of $E_T\!\!\!\!\!\!/~~$ can not change the reach significantly. This feature is in contradiction 
with most of the supersymmetric models. In a recent study \cite{Baer:2010tk}, 
it was shown in the context of minimal supergravity model that use of $E_T\!\!\!\!\!\!/~~$ 
cut will enhance the LHC reach significantly. However, in our case, if we do not use the $E_T\!\!\!\!\!\!/~~$ cuts, the reach of $R^{-1}$ slightly reduced except in the tri-lepton channel (see Fig. \ref{fig:reach_without}).

Here also, for low integrated luminosity, OSD and tri-lepton channels are most promising. However,
for high integrated luminosity, tri-lepton is the most promising channel because of the 
low SM background. As for example, with $2~{\rm fb}^{-1}$ integrated 
luminosity, mUED parameter space can be probed upto $R^{-1}=714~(675)$ GeV for $\Lambda R=40~(10)$ 
in the tri-lepton channel. However, in the SSD and OSD channel the discovery reach is 
$R^{-1}=683~(653)~{\rm and}~650~(650)$ GeV respectively for  $\Lambda R=40~(10)$. As we increase 
$R^{-1}$, the total pair production cross-section of $Q_1~{\rm and}~g_1$ decreases 
(see Fig. \ref{fig:cross_7_14}) and hence, the signal cross-section decreases. In defining the observability of 
the signal, we have demanded that $\sigma_S>\sigma_B/5$. The SM background for the OSD channel is 
quite large. Therefore, we can not probe $R^{-1}$ beyond $650$ GeV in the OSD channel by increasing 
integrated luminosity.  

Here we do not show the discovery 
reach in the 4 lepton channel separately because the 4 lepton cross section is much smaller than other 
three processes considered here.  As for example, we obtain 4-lepton cross-section 
(with the $E_T\!\!\!\!\!\!/~~$ cuts) of 3.21 (1.69) fb for $R^{-1}=400~(600)$ GeV and 
$\Lambda R=20$. Therefore, we have not found atleast 5 events in most of the parameter space 
points. It is possible to combine the significance of these channels but no such attempt has made in our
paper.    

\section{Conclusion}
In this paper we have investigated the early discovery signals of minimal UED model at 
the Large Hadron Collider with center of mass energy 7 TeV and integrated luminosity 1fb$^{-1}$.
We have considered signals with two or more leptons in the final state and we have calculated the
discovery reach in each of the final states. The results of our 
work indicates that LHC can discover the signals of mUED if $R^{-1}$ is less than 700 GeV. The 
opposite sign di-lepton channel is the most promising one. We have seen that the inclusion of missing 
energy does not change the discovery limit drastically. This is due to the fact that the 
$E_T\!\!\!\!\!\!/~~$ distribution of mUED model has the similar shape as the SM backgrounds. It should 
be stressed that our cuts are very simple and reach may be improved by imposing more specific 
cuts. In summary, we can say that early LHC data has substantial reach within the mUED parameter 
space.   

{\bf Acknowledgement}: The authors would like to thank the organizers of WHEPP-XI (Ahmedabad, 2010) where 
some of this work was done. BB thanks Sreerup Raychaudhuri and Anirban Kundu for many useful discussions. 
KG acknowledges the support available from the Department of Atomic Energy, Government of India, for the Regional Centre for Accelerator-based Particle Physics (RECAPP), Harish-Chandra Research Institute. Computational work for this study was partially carried out at the cluster computing facility in the Harish-Chandra Research Institute (http://cluster.hri.res.in).

\end{document}